\documentclass[aps,pre,twocolumn,nofootinbib,longbibliography,tightenlines,superscriptaddress,nopacs,amsmath,amssymb,final,letterpaper]{revtex4-1}

\usepackage{graphicx}
\usepackage{floatrow}
\usepackage[caption=false,position=top]{subfig}
\usepackage{dcolumn}
\usepackage{bm}
\usepackage{xcolor}
\usepackage{amsmath,amssymb,mathrsfs,physics,tikz}
\newcommand{\RN}[1]{%
\textup{\uppercase\expandafter{\romannumeral#1}}%
}
\usepackage{hyperref}
\hypersetup{
    colorlinks=true,
    linkcolor=blue,
    filecolor=magenta,      
    urlcolor=cyan
    }

\usepackage{float}

\begin{document}

\setlength{\abovedisplayskip}{3pt}
\setlength{\belowdisplayskip}{6pt}
\setlength{\abovecaptionskip}{3pt}
\setlength{\belowcaptionskip}{6pt}

\newcommand{\argmin}{\operatornamewithlimits{argmin}}
\newcommand{\argmax}{\operatornamewithlimits{argmin}}

\title{Characterizing network circuity among heterogeneous urban amenities}

\author{Bibandhan \surname{Poudyal}}
\affiliation{Department of Physics \& Astronomy, University of Rochester, Rochester, NY 14607, USA}

\author{Gourab \surname{Ghoshal}}
\email{gghoshal@pas.rochester.edu}
\affiliation{Department of Physics \& Astronomy, University of Rochester, Rochester, NY 14607, USA}

\author{Alec \surname{Kirkley}}
\email{alec.w.kirkley@gmail.com}
\affiliation{Institute of Data Science, University of Hong Kong, Hong Kong}
\affiliation{Department of Urban Planning and Design, University of Hong Kong, Hong Kong}
\affiliation{Urban Systems Institute, University of Hong Kong, Hong Kong}


\begin{abstract}
The spatial configuration of urban amenities and the streets connecting them collectively provide the structural backbone of a city, influencing its accessibility, vitality, and ultimately the well-being of its residents. Most accessibility measures focus on the proximity of amenities in space or along transportation networks, resulting in metrics largely determined by urban density alone. These measures are unable to gauge how efficiently street networks can navigate between amenities, since they neglect the circuity component of accessibility. Existing measures also often require ad hoc modeling choices, making them less flexible for different applications and difficult to apply in cross-sectional analyses. Here we develop a simple, principled, and flexible measure to characterize the circuity of accessibility among heterogeneous amenities in a city, which we call the pairwise circuity (PC). The PC quantifies the excess travel distance incurred when using the street network to route between a pair of amenity types, summarizing both spatial and topological correlations among amenities. Measures developed using our framework exhibit significant statistical associations with a variety of urban prosperity and accessibility indicators when compared to an appropriate null model, and we find a clear separation in the PC values of cities according to development level and geographic region.   
\end{abstract}

                              
\maketitle

\section{Introduction}

The layout of urban amenities and street network infrastructure in a city form its structural foundation, facilitating human mobility, the exchange of goods, services and ideas, and the visual character of the city \cite{barthelemy2016structure,bettencourt21,boeing2018measuring, Pan2013}. This urban structure in turn exerts a profound influence on the well-being and socioeconomic prosperity of urban residents \cite{guite2006impact,witten2003quality,king2015disadvantaged,bereitschaft2017equity, Bassolas2019}. Given the wealth of newly available data providing high resolution information about a wide range of urban amenities and infrastructure, there is great interest from researchers and government entities in identifying urban indices that succinctly summarize this data and separate structure from noise \cite{kandt2021smart,boeing2020multi,kitchin2015knowing}. 

In her early pioneering work ``The Death and Life of Great American Cities'' \cite{jacobs2016death}, Jane Jacobs proposed that a mix of land uses, small block sizes, coexistence between old and new buildings, and high developmental density are the four major factors that determine the ``vitality'' of a city---broadly, the capability for the city to promote a range of activities among diverse populations throughout the day, enhancing liveability and deterring crime and urban decay. Underlying this characterization of urban vitality is the concept of heterogeneity---how evenly different types of activities and amenities are distributed---and implicit in the above criteria are both spatial and topological notions of heterogeneity among urban amenities. For example, the concepts of land use mix and building coexistence encompass the (spatial) distance and (topological) adjacency between land partitions of different uses and buildings of different ages respectively \cite{eom2020land,lu2019impacts}. Jacobs additionally states that accessibility to urban amenities---particularly through walking, bicycling, or public transport---is a critical factor that facilitates vitality \cite{sung2015residential}. The accessibility of amenities in a city is naturally influenced by both their distance and adjacency along the infrastructure network, invoking the notions of spatial and topological proximity as well as circuity.  

Existing research has noted the importance of accounting for both spatial and topological correlations to understand a diverse range of urban phenomena, including housing prices \cite{yuan2012discovering}, the emergence of a city center \cite{zhu2017building}, human mobility preferences \cite{lee2017morphology}, and urban spatial segregation \cite{kirkley2022spatial}. For example, in \cite{kirkley2020information} a new framework to understand correlations in spatial socioeconomic data was proposed which endows the network of spatial adjacencies among government-designated regions with distance weights reflecting the Jensen-Shannon divergence between the data distributions associated with the regions. And in \cite{arcaute2016cities}, regions obtained as connected components in street network percolation processes are identified with natural, socioeconomic, and administrative boundaries in Britain.   

Several studies have also suggested the importance of urban amenity configuration to vitality and accessibility, focusing either on the spatial proximity of amenities \cite{smoyer2004spatial,das2021access,burdziej2019using,smith2010neighbourhood,hewko2002measuring,tannier2012spatial,yuan2020amenity,logan2021measuring,talen2022can} or their adjacency along street infrastructure \cite{sung2015residential,fang2021spatial,sulis2018using,sulis2019measuring,kim2020urban,zhang2021can,xia2020analyzing}. Existing work that has focused on both the spatial and topological facets of accessibility or vitality have largely been restricted to a specific class of urban amenities such as transportation facilities \cite{olszewski2005using,rastogi2003defining,geurs2015accessibility,landex2006examining,garcia2018analysing}, healthcare facilities \cite{boscoe2012nationwide,bell201310,henry2013geographic}, educational institutions \cite{yenisetty2020spatial}, entertainment \cite{he2019spatial}, and greenspace \cite{teeuwen2023measuring}. A few recent studies have aimed at capturing urban vitality or accessibility comprehensively using aggregate indices that capture multiple spatial and topological factors simultaneously \cite{zhao2018towards,tu2020portraying,li2021multidimensional,zhang2021analysis}. However, these studies either combine existing measures in a complex ad hoc manner or identify combinations of different measures empirically by optimizing the correlation with indicators such as social media activity or population in-flows.

In a recent paper, Bassolas and Nicosia \cite{bassolas2021first} develop a framework to measure the structural correlations and heterogeneity in a range of complex systems that relies on the concept of the mean first passage time for random walks on networks \cite{redner2001guide}. This method takes as input a network with metadata categorizing each node into one of a small number of classes and computes the class mean first passage time (CMFPT) between each pair of node classes, defined as the expected number of steps it takes for a random walk along the network to reach one class when starting at the other. This framework can be used to capture the topological nature of heterogeneity and correlations in complex systems, providing new insights into a variety of phenomena including the spread of epidemics and segregation \cite{bassolas2021diffusion}. However, as this method is aimed at capturing correlations in general complex networks, it is formulated without an explicit dependence on spatial density, so cannot be directly applied to understand urban accessibility and vitality. Additionally, although it allows for elegant analytical expressions, the dynamical formulation in \cite{bassolas2021first} based on random walks is not clearly tied to real human movement along street networks, which provides the topological proximity of interest for cities. Although they are influenced by navigation heuristics and the surrounding built environment \cite{bongiorno2021vector,miranda2021desirable}, human trajectories in street networks are much more highly correlated with the shortest paths in the network than completely random walks. 

The proximity of amenities (1) in space---reflected by spatial density---and (2) along networks representing various transport modes forms the foundation of most existing work on accessibility \cite{smoyer2004spatial,das2021access,burdziej2019using,smith2010neighbourhood,hewko2002measuring,tannier2012spatial,yuan2020amenity,logan2021measuring,talen2022can,sung2015residential,fang2021spatial,sulis2018using,sulis2019measuring,kim2020urban,zhang2021can,xia2020analyzing}. These two factors are of critical importance for understanding accessibility, but they do not tell the whole story. From a network science perspective, it is of great interest to understand how we can design cost effective networks that facilitate efficient transportation---in other words, direct (not circuitous) access---among existing points of interest \cite{barthelemy2011spatial}. Measures that solely capture the spatial density of amenities or their proximity along these networks provide little use for this task, which can be seen in the following example. Consider a city which is choosing where to supplement their sidewalk network in order to provide better pedestrian access between public transport stations and offices. The natural locations to choose are ones in which the new sidewalks would reduce the total travel distance the most, which may or may not correspond to the locations in which the distance along the existing sidewalk is the greatest. If a suburban area has lots of office buildings located at a 30 minute walk from the nearest public transport stop, but the sidewalk route from the offices to the transport stop is very direct (straight), then there is little one can do to improve the sidewalk network to increase its efficiency for accessing the public transport stops. (One could, however, add a public transport stop closer to the offices.) On the other hand, in a dense urban environment, it may be beneficial to add a new shortcut between offices and a public transport stop even if the offices are already only a 10 minute walk away from the stop, if it will cut the journey to 3 minutes. In the first case, the existing sidewalk network provided very direct access from offices to public transportation (i.e., had a low level of circuity among the two amenities), while in the second case the network had very indirect access from offices to public transportation (i.e., had a high level of circuity among the two amenities).

In this paper we develop a simple, principled, and flexible measure to characterize the circuity of routes between amenities along a city's street infrastructure, which captures both spatial and topological notions of heterogeneity and proximity among the amenities. Our measure is inspired by the CMFPT \cite{bassolas2021first} and the detour factor \cite{barthelemy2011spatial} (also known as the route factor, circuity, or directedness), defined as the ratio of the network distance to the Euclidean (``crow-fly'') distance between points of interest in a street network with distance-weighted edges. We call our measure the pairwise circuity (PC), which is computed as the average excess travel distance incurred when using the
street network to route between a given pair of amenity classes along the shortest network path. By averaging the pairwise circuity values over multiple amenity classes, we obtain aggregate measures of the circuity among urban amenities that can be interpreted as the expected value of the pairwise circuity when starting at a randomly chosen amenity. Compared to a null model where the amenity classes are randomly shuffled while fixing the class frequencies, we find that both the aggregated PC measures and PC values for individual amenity pairs exhibit statistically significant correlations with a range of urban prosperity and accessibility indicators across cities worldwide. We also find a clear ordering in the distributions of PC values over groups of cities determined by economic development and geographical region, results that are also highly robust relative to the null model. Our measure can succinctly summarize the density and correlations among different classes of urban amenities in a multifaceted manner as well as provide a simple measure of the circuity of amenity accessibility to complement existing methods that assess urban vitality and accessibility from a structural perspective.

\section{Methods}

\subsection{Data Description}
\label{sec:data}

Using the OSMnx Python package \cite{boeing2017osmnx} which calls the OpenStreetMap API \cite{OpenStreetMap}, we collected open source data on street geometries and amenity locations for 371 cities worldwide (Fig. S1). Since we are particularly interested in the directness of accessibility to amenities through walking, bicycling, and public transport, we used the pedestrian layer of each street network in our study. We choose the default OSM amenity categories as amenity classes in order to avoid imposing our own beliefs about the amenity categories and take an accepted classification used in previous studies \cite{chen2021delineating,hu2016impacts,s2022alf,papadakis2019function,barbosa2021uncovering}. See Table~\ref{tab-amenities} for details on the amenity classes and their typical frequencies across the cities for which data was collected. 

We remain agnostic about the choice of amenity classes and include all amenity classes in our study for multiple reasons. First, the aim of this paper is to present the new Pairwise Circuity (PC) framework, which can be adapted for any amenity classification scheme depending on the application of interest. For example, for tourism-centered analyses it may be of interest to focus only on the circuity among different `entertainment' (museums, cinemas etc) and `sustenance' amenities (bar, cafe, restaurant, etc). In this case, one may want to ignore all other amenity classes but use a more fine-grained classification scheme than `entertainment' and `sustenance', since the specific type of venues within these categories will be of high importance. Second, the amenity pairs that contribute to one's perceived urban vibrancy and accessibility as a whole can vary highly based on personal circumstances: elderly or immunocompromised individuals may place a higher relative importance on the accessibility of healthcare-related amenities than younger demographics. Meanwhile, single parents may place higher relative importance on accessibility among amenities that aid with running day-to-day errands (e.g. `financial', `waste', and `public service' amenities). Finally, the contributions of different amenity types to urban vitality may vary substantially from city to city. Circuity among `entertainment' amenities (e.g. casinos) might contribute more towards the overall urban vibrancy in Las Vegas than in New York City, where `transportation' amenities may play a comparatively large role in vibrancy due to density and vehicular congestion.

For the subset of the cities included in the Jones Lang LaSalle (JLL) report on global cities \cite{JLL}, we use the available classifications of city development level assigned according to each city's real estate, corporate occupier base, and commercial stock. (These classifications were also used for the analysis in \cite{mimar2022connecting}, where they are described in further detail.) We aggregated the cities under the JLL categories ``Super'', ``Matured'', and ``Transitional'' into a single ``Matured'' category, and aggregated the categories ``Developing'', ``Early Growth'' and ``Nascent'' into the ``Developing'' category. 

Furthermore, we classified all 371 cities into two regions: (1) North America, Western Europe and Australia/New Zealand (195 cities); (2) Africa, Eastern Europe, South America, and Asia (176 cities). This dichotomy roughly corresponds with the Global North/South divide as well as the division between developed and developing countries according to International Monetary Fund \cite{IMF}.  The distribution of cities based on these two different types of classification is shown in Fig. S2. In Figure S3 we plot the city diameter, average shortest path length and the distribution of amenities split by the classifications. We find that Mature cities and Region 1 cities are more compact given the lower diameter and shortest path lengths. In contrast, the distribution of amenities across the classifications is more or less identical.

\begin{table}[h]
\renewcommand{\arraystretch}{1.5}
\begin{tabular}{ | p{4.3cm}| p{2.10cm}| p{1.7cm}|}
 \hline
 \multicolumn{3}{|c|}{\textbf{Amenity Classes}} \\
 \hline
\textbf{Tags} & \textbf{Class} & \textbf{Frequency}\\
 \hline
 bar, cafe, restaurant, etc & Sustenance & 1604 $\pm$ 124
 \\
 \hline
college, school, university, etc & Education & 599 $\pm$ 40
\\
\hline
 bus station, fuel, parking, etc &  Transportation & 3417 $\pm$ 313
 \\
 \hline
 atm, bank, money transfer, etc &  Financial & 335 $\pm$ 26
 \\
 \hline
 clinic, hospital, pharmacy, etc & Health & 529 $\pm$ 43
 \\
 \hline
arts center, casino, cinema, etc & Entertainment & 348 $\pm$ 23
\\
\hline
police station, post box, etc &  Public Service & 310 $\pm$ 36
 \\
 \hline
 bbq, bench, locker, etc &  Facilities & 1333 $\pm$ 141
 \\
 \hline
 recycling, waste basket, grit bin, etc &  Waste & 661 $\pm$ 85
 \\
 \hline
apartments, dormitory, childcare, etc &  Others & 549 $\pm$ 37
 \\
\hline
\end{tabular}
\caption{Amenity classifications provided by OpenStreetMap \cite{wiki}, which are used for the example applications in Sec.~\ref{sec:results}. The mean frequency and corresponding standard error across the 371 cities studied are listed next to each amenity class.} 
\label{tab-amenities}
\end{table}

As amenity accessibility and diversity is highly correlated with various aspects of socioeconomic and environmental well-being in cities \cite{witten2003quality}, we also collected a variety of prosperity indicators across multiple socioeconomic and environmental facets with which we compare our measures. We collected data from several UN sources on the Gini coefficient, Internet access rate, public transportation access, GDP per capita, quality of life index, poverty rate, infrastructure index, Local Online Service Index (LOSI) \cite{LOSI}, public space allocation, and public space access for cities in our dataset. We also compare our measures with the Walk Score and Bike Score indices \cite{WalkScore} which are widely used measures of the walkability and bikeability of cities across North America and Western Europe \cite{carr2010walk,duncan2011validation}. The availability of data was different for different sources. For prosperity metrics from UN sources, the number of cities with available data varies from 49 (Gini Coefficient) to 153 (Transportation Access), whereas the Walk Score and Bike Score data was available for 114 US and Canadian cities as well as London (115 cities in total). All measures analyzed, along with the number of cities for which data was available, are detailed in Table S1 in the Supplementary Material.

\subsection{Mathematical Formalism}

Suppose there are $N$ total amenities (alternatively, facilities or points of interest) indexed by $i=1,...,N$, with $(x,y)$ coordinates $\{(x_i,y_i)\}_{i=1}^{N}$. Each amenity $i$ is grouped into one of $C$ amenity classes such that $c_i\in \{1,...,C\}$ is the amenity class of point $i$. The amenity class $c_i$ gives a generic categorization of the type of service provided at the point of interest $i$. For example, amenities $i$ such as cinemas or museums may be classified with $c_i=\text{``entertainment''}$ to indicate their broad categorization as amenities aimed at leisure entertainment \cite{he2019spatial}. We let $n_c$ denote the number of amenities in class $c$, so that $\sum_{c=1}^{C}n_c=N$.

The classification scheme mapping amenities $i$ to classes $c_i$ will in general have an impact on the results of our method, so constitutes an important choice for a practitioner using the method. As our method is applicable to any partition of the amenities into classes, the amenity classification scheme can be used to reflect the distinctions among amenities relevant to a particular application (e.g. bus, train, and subway stations in a transportation-focused analysis; or commercial and residential properties in a zoning-oriented study). In the example applications we present here, we use the pre-defined classes of amenities provided by OpenStreetMap \cite{wiki}, from which the data were collected (see Table~\ref{tab-amenities} for a summary and Sec.~\ref{sec:data} for details). 

Along with the $N$ amenities distributed in space, there is a street network $G=(V,E,W)$ embedded in space that is in the primal representation \cite{barthelemy2011spatial}. In this representation, the nodes $v\in V$ represent intersections, an edge $(u,v)\in E$ exists between two nodes $u,v\in V$ if and only if there is a street segment directly connecting their corresponding intersections, and the edge $(u,v)$ is endowed with the weight $w_{uv}$ representing the distance of its corresponding street segment. 

\begin{figure*}[t]
\includegraphics[width=13cm,angle=270]{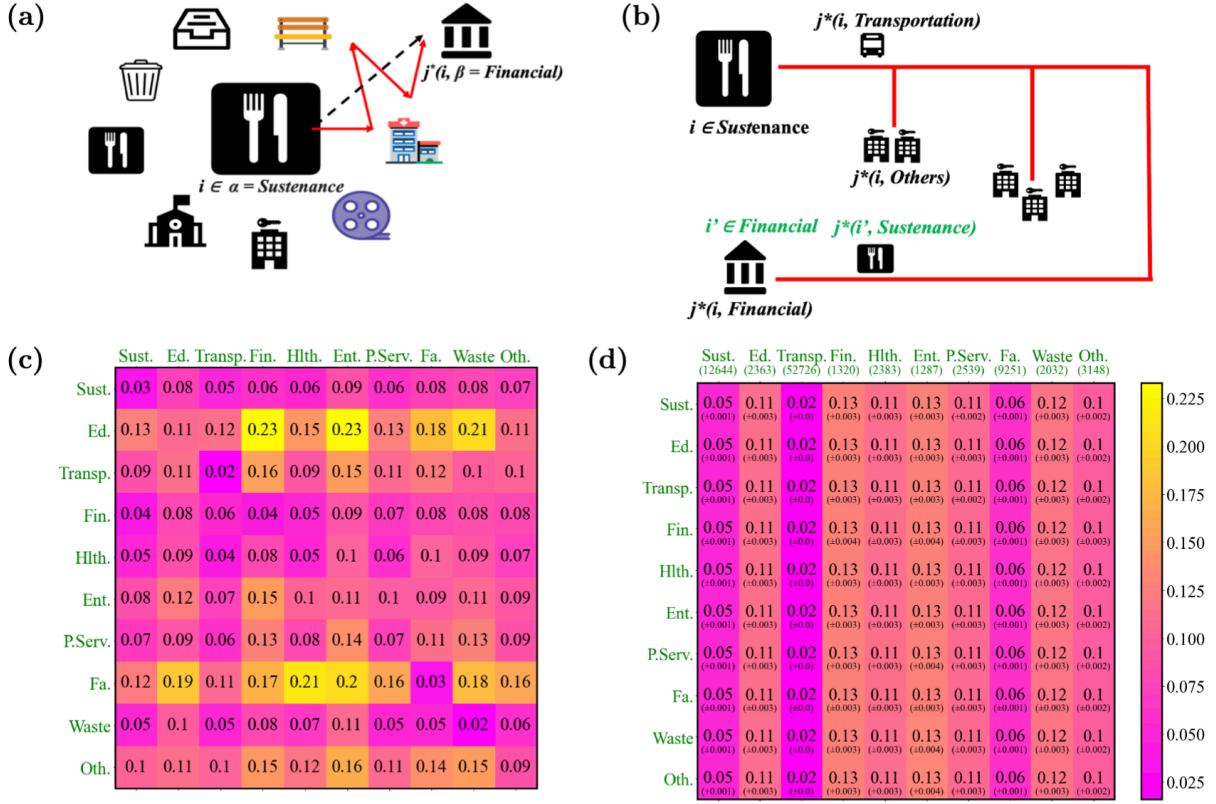}
    \caption{\textbf{The Pairwise Circuity (PC) measure.} \textbf{(a)} For each origin amenity class $\alpha$ (here, ``Sustenance'') and destination amenity class $\beta$ (here, ``Financial''), the Pairwise Circuity $\text{PC}_{\alpha\beta}$ (Eq.~\ref{Pairwise Circuity}) quantifies the average excess travel distance required to route from a node in class $\alpha$ to the nearest node in class $\beta$. Each amenity $i$ in class $\alpha$ contributes to this average a term $d_G(i,j^\ast(i,\beta))-d_E(i,j^\ast(i,\beta))$, which gives the difference between the street network distance $d_G$ (solid red line) and the crow-fly distance $d_E$ (dotted black line) between amenity $i$ and the nearest amenity in class $\beta$, $j^\ast(i,\beta)$. \textbf{(b)} If we start at the amenity of interest (amenity $i$ in the ``Sustenance'' class) and the nearest amenity of a particular class is accessible via a straight street segment (here, ``Transportation''), the contribution of this amenity pair to the Pairwise Circuity is $0$. On the other hand, if the street distance to the closest amenity in the class is large compared to the crow-fly distance, then the Pairwise Circuity contribution of this amenity pair is large and provides evidence of a lack of direct accessibility between the two amenities along the network. The Pairwise Circuity between two amenities can be asymmetric (here, ``Sustenance'' to ``Financial'' vs ``Financial'' to ``Sustenance'', with corresponding variables in black and green respectively). \textbf{(c)} Pairwise Circuity matrix of New York City (values in km). \textbf{(d)} Pairwise Circuity matrix of New York City, after shuffling the class labels of the amenities while maintaining the overall class frequencies. Values indicate the mean and standard error over 100 realizations of amenity shuffling. The relative Pairwise Circuity values are not preserved under this null model, indicating that the Pairwise Circuity is reflecting correlations in amenity locations along the street network beyond the information provided by the spatial density and relative frequencies of the amenities.}
    \label{Description}
\end{figure*}

Using the amenities and street network, we can define the crow-fly distance $d_E(i,j)$ between two amenities $i,j$ as the usual Euclidean distance, which is valid at small length scales. We can also define the network distance $d_G(i,j)$ between the two amenities as the distance along the shortest street network path connecting $i$ and $j$. To get this path length, we project each of $i$ and $j$ onto the nearest points on the street network (which may or may not be vertices in $V$) and compute the shortest path between the two projected amenities along the street network using Dijkstra's algorithm \cite{dijkstra2022note}, accounting for the distance required to project the points onto the network. We note that, by construction, $d_E(i,j) \leq d_G(i,j)$, since routing along the street network will always incur some additional cost due to its limited coverage of space. 

In principle, one can augment the network distance $d_G$ by incorporating the cost of various inconveniences such as angular deviation, slope, or congestion \cite{cooper2020sdna}. Along similar lines, one could transform both the distance measures $d_E$ and $d_G$ into travel times. In these cases, our measure would be interpreted in units of cost or time rather than distance. Furthermore, to compute $d_G(i,j)$ one could use a different notion of network distance between amenities $i$ and $j$ such as random walk-based measures \cite{redner2001guide,hanna2021random}. Such choices would result in different interpretations of our measures, and may be more relevant for certain contexts. For example, travel time may be a more relevant indicator than travel distance when analyzing the network circuity for car-based routing, where there may be higher levels of variability between distance covered and travel time due to vehicular congestion and speed limits, which are less relevant for pedestrian travel. However, such extensions require specialized data and/or expensive computer simulations which inhibit scaling our analysis to a large corpus of global cities, so we do not explore these options here.

\subsection{Pairwise Circuity}
\label{sec:Pairwise Circuity}

In order to capture both spatial and topological notions of the circuity of accessibility among amenities as well as the heterogeneity of these amenities, we can compare the distances $d_E(i,j)$ and $d_G(i,j)$ between pairs of amenities $i,j$ that fall into different classes $c_i,c_j$. Intuitively, if $d_G(i,j)\gg d_E(i,j)$ for many pairs $i,j$ in classes $r,s$ respectively, then the street infrastructure of the city is not providing direct access between amenities of types $r$ and $s$, which may have adverse effects on the accessibility of the city from the structural perspective. On the other hand, if $d_G(i,j)\approx d_E(i,j)$ for most pairs $i,j$ with classes $r,s$, then we do not incur much extra travel cost in routing between the amenity classes $r$ and $s$. 

There are a few choices for quantifying the deviation of $d_G(i,j)$ and $d_E(i,j)$. Perhaps the most natural choice is the detour factor \cite{barthelemy2011spatial,kirkley2018betweenness} (also known as the route factor, circuity, or directedness), which measures the ratio $d_G(i,j)/d_E(i,j)$. The detour factor can be averaged over all pairs $i,j$ to get an idea of the typical relative excess travel cost incurred by traveling between a pair of nodes. At first glance, the detour factor appears to have the advantage of being scale-independent, since it divides out spatial scale factors in the numerator and denominator. However, in practice the detour factor tends to decrease as trips become longer \cite{yang2018universal}, due to the relatively straight nature of long street routes. Therefore, the detour factor makes it appear as if short paths are very inconvenient, when in practice the additional travel distance incurred may be negligible. Moreover, the absolute spatial density of development is a critically important quantity that influences the well-being of a city \cite{jacobs2016death}, so by dividing out the spatial distance in the detour factor expression one ignores this important factor. 

Since the travel distance (or cost, time, etc, depending on the definitions of $d_E$ and $d_G$) itself is a more relevant quantity than the relative deviation quantified by the detour factor, for our measure we opt for the difference $d_G(i,j)-d_E(i,j)$ as the quantity of interest quantifying meaningful deviations due to street network inefficiencies. This quantity can be directly interpreted as the excess travel distance incurred when routing from $i$ to $j$ due to the street network connectivity and geometry. 

By aggregating the excess travel distance $d_G(i,j)-d_E(i,j)$ over pairs $i,j$ with $c_i=r,~c_j=s$, we can identify amenity classes $r,s$ that have high directness of accessibility to each other (low excess travel distances), and vice versa. However, it is not realistic to simply take the average $d_G(i,j)-d_E(i,j)$ over all pairs $i,j$ such that $c_i=r,~c_j=s$, since---under the assumption of equivalence among the amenities within a given class---it is unlikely that any rational agent would choose to route to an amenity that is much farther than a closer alternative in the same class \cite{prato2009route}. 

In the case of additional information about routing preferences, one can employ more sophisticated route choice models that provide non-zero weights to different alternatives within the same amenity class \cite{prato2009route}. These models will increase the computational burden of the method, but in principle should not change the results substantially so long as the closest alternative is given the highest weight. Additionally, with population density information, one could weight the excess travel distance between a pair of amenities using gravity-type models of travel demand \cite{barbosa2018human} to obtain an expected net incurred cost across all travellers. With the goal of presenting a purely structural measure capturing the circuity of accessibility among heterogeneous urban amenities that incorporates minimal assumptions, we do not explore these extensions in this paper. 

With this in mind, in the absence of real demand data which is hard to obtain, we will assume that agents routing along the network are primarily interested in the closest amenity of each class. This allows us to define the \emph{Pairwise Circuity} (PC) between the amenity classes $\alpha$ and $\beta$ as
\begin{align}\label{Pairwise Circuity}
\text{PC}_{\alpha\beta} &= \frac{1}{n_\alpha}\sum_{i=1}^{N}\Big[d_G(i,j^\ast(i,\beta)) -d_E(i,j^\ast(i,\beta))\Big]\delta_{c_i,\alpha},       
\end{align}
where $\delta_{c_i,\alpha}$ is the Kronecker delta function restricting us to origin nodes $i$ that are in class $\alpha$, $n_\alpha$ is the number of POIs belonging to amenity class $\alpha$, and
\begin{align}
j^\ast(i,\beta )= \argmin_{j:c_j=\beta}\{d_E(i,j)\}   
\end{align}
is the amenity $j$ in class $\beta$ that is closest to amenity $i$ in space. We can see that in general $\text{PC}_{\alpha\beta} \neq \text{PC}_{\beta\alpha}$, indicating that amenity class $\beta$ may be more directly accessible along the streets to nodes in class $\alpha$ than amenity class $\alpha$ is to nodes in class $\beta$. (The same asymmetry can be found in the CMFPT of \cite{bassolas2021first}.) A schematic illustrating the Pairwise Circuity measure is shown in Fig.~\ref{Description}a,b. 

The Pairwise Circuity information for a city can be summarized in a single (asymmetric) $C\times C$ matrix with values $\text{PC}_{\alpha\beta}$ for $\alpha=1,...,C$ and $\beta=1,...,C$. The diagonal values of this matrix, $\text{PC}_{\alpha\alpha}$, correspond to the circuity of accessibility among amenities within the same class, while off-diagonals $\text{PC}_{\alpha\beta}$ correspond to the circuity of accessibility among amenities of different classes. An example for New York City is shown in Fig.~\ref{Description}c, where the matrix elements are in units of km. The figure indicates that the diagonals of the Pairwise Circuity matrix are in general lower than the off-diagonals, reflecting the agglomerative nature of many amenities \cite{rosenthal2004evidence}, although there are heterogeneities in the level of clustering signalled by varying magnitudes in the diagonal elements. For instance, we find that transportation facilities are the most directly accessible from each other, with their network distance differing by only 20 meters on average from their Euclidean distance, while the length of streets connecting entertainment facilities are on average 110 meters greater than the corresponding crow-fly distances.  

\begin{table*}[t]
\renewcommand{\arraystretch}{1.5}
\begin{tabular}{ | p{2.75cm}| p{4.85cm}| p{9.9cm}|}
 \hline
 \multicolumn{3}{|c|}{\textbf{Amenity Circuity Measures}} \\
 \hline
\textbf{Measure} & \textbf{Definition} & \textbf{Interpretation}\\
 \hline
 Pairwise Circuity & Eq.~\ref{Pairwise Circuity} & Expected excess travel distance along the street network (e.g. deviation from crow-fly distance) to reach the nearest amenity of the specific class $\beta$, starting at a randomly chosen amenity in class $\alpha$. 
\\
\hline
Marginal Circuity & $\text{MC}_\alpha = \frac{1}{C-1}\sum_{\beta\neq\alpha } \text{PC}_{\alpha\beta}$ &
Expected excess travel distance along the street network to reach the nearest amenity within a randomly chosen class $c\neq \alpha$, starting at a randomly chosen amenity in class $\alpha$.
 \\
 \hline
Average Circuity & $\text{AC} = \frac{1}{C-1}\sum_{\alpha=1}^{C}\frac{n_\alpha}{N}\sum_{\beta\neq\alpha } \text{PC}_{\alpha\beta}$ & Expected excess travel distance along the street network starting at a randomly chosen amenity and ending at the nearest amenity within a randomly chosen class different from the starting class.
 \\
\hline
\end{tabular}
\caption{Definitions and interpretations of the Pairwise Circuity-based measures used in this paper.} 
\label{tab-definitions}
\end{table*}

By its definition, the Pairwise Circuity matrix is affected by spatial density---as intended, since this is a key component of accessibility and vitality. However, in the experiments in Sec.~\ref{sec:results} we are interested in identifying how the correlations among amenities in global cities impacts Pairwise Circuity results while controlling for the variability in the spatial density as well as frequency of different amenity classes across the cities, as these may be affected by heterogeneity across different regions due to OSM sampling coverage, cultural differences, etc. We therefore compare any Pairwise Circuity results we obtain for real data with the results of simulations from a null model that preserves the overall spatial density of amenities and the frequency of each amenity class but destroys the correlations among the amenities in space and along the street network.

In the null model we consider, the positions of the amenities remain fixed but the tags of the amenities (e.g. ``bar'', ``atm'', ``waste basket'') are shuffled uniformly at random. Each tag is associated with an amenity class categorizing the amenity type (see Table~\ref{tab-amenities}), and shuffling the tags uniformly at random is equivalent to shuffling the amenity class labels uniformly at random while preserving their frequencies across the city. By examining a the average Pairwise Circuity matrix over many realizations of this null model for New York City (Fig.~\ref{Description}d), we can see that spatial density and amenity frequency alone cannot explain the empirically observed Pairwise Circuity values (Fig.~\ref{Description}c), since the relative values are not preserved. Instead, we can see that in the null model, the Pairwise Circuity values are determined entirely by the frequency of the destination amenity class (i.e. the columns of the Pairwise Circuity matrix are constant). This is because, regardless of our starting point, the distance to the nearest amenity of a certain class in the null model will be determined solely by the frequency of the amenity class---more frequent amenity classes will on average be closer---since all tags are being shuffled uniformly at random. The frequency of the origin amenity class does not matter in this case because over all possible label permutations each class is equivalent in terms of its expected proximity to the closest node in each other class.   

From the Pairwise Circuity matrix one can extract a number of useful aggregate measures, a few of which we explore in the next section. 

\subsection{Aggregate Pairwise Circuity Measures}

The Pairwise Circuity matrix contains information about all pairs of different amenity classes in a city, allowing us to understand the circuity of accessibility among specific amenity types. However, for large-scale comparisons across cities it is useful to extract a smaller set of values from this matrix that capture more aggregated notions of circuity. 

\begin{figure*}[t]
    \includegraphics[width=12cm]{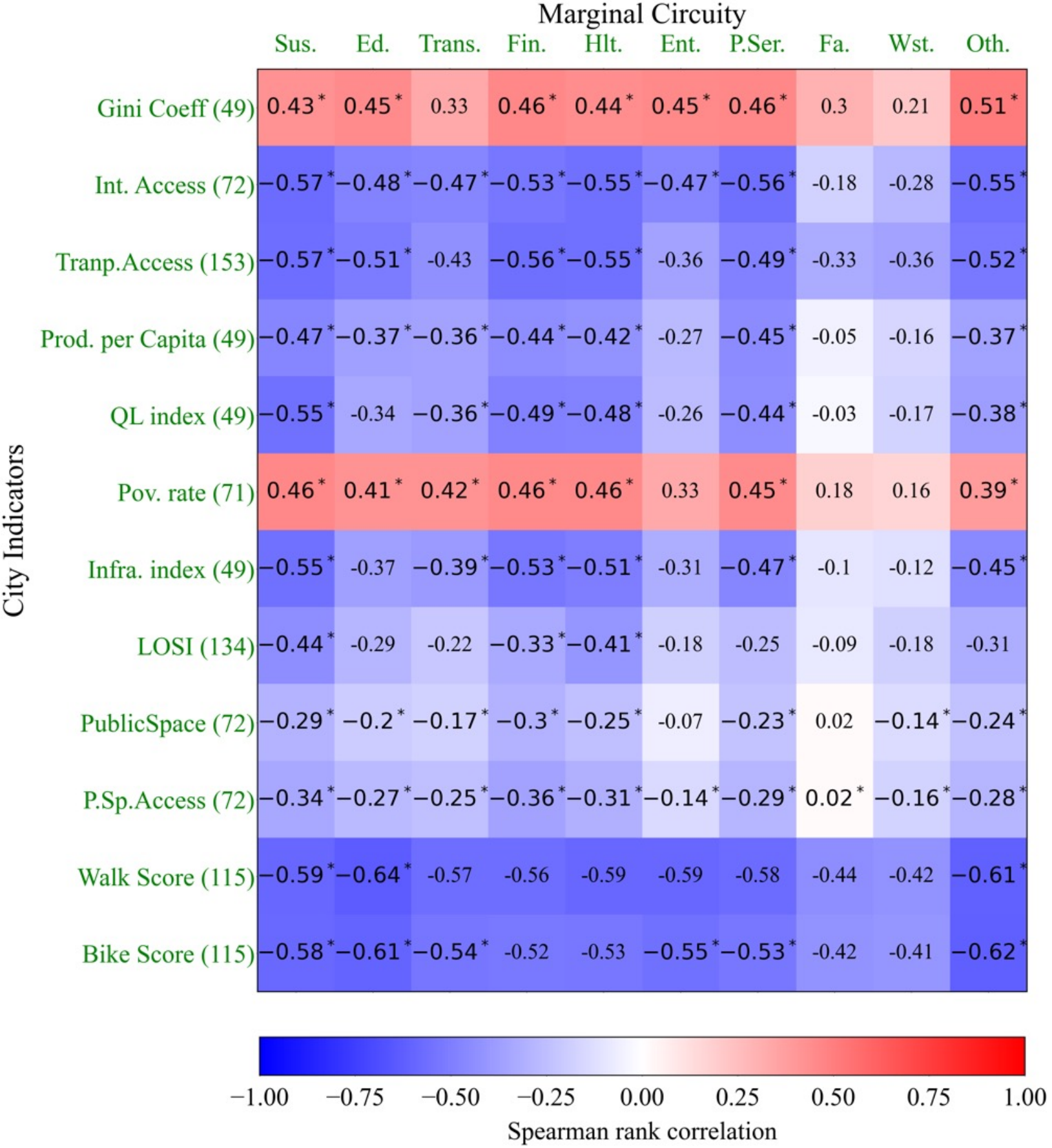}
      \caption{\textbf{Associations among Marginal Circuity values and prosperity/accessibility indicators across global cities.} The Marginal Circuity (Eq.~\ref{eq:Marginal Circuity}) was computed for all cities with available data for each indicator (number indicated in parentheses) and all amenity classes. Superscript $*$ indicates a permutation test $p$-value of less than $0.05$ over 100 draws from the null model where the class labels of the amenities are shuffled while maintaining the overall class frequencies. The number in the bracket next to the indicators represents the number of cities for which indicator data was available. The urban indicators are defined in Table S1.}
    \label{Correlation}
\end{figure*}

A natural measure we can compute from the Pairwise Circuity matrix is its average, which gives an overall length scale of expected excess travel distance between amenity classes. However, there are a couple of issues with the interpretability of a simple average over Pairwise Circuity matrix entries. Firstly, it treats all amenity classes on an equal footing, while some amenity classes may be much more frequent than others. Secondly, it includes competing contributions from the diagonals---which characterize the within-class accessibility and will favor amenity classes with high levels of agglomeration---and the off-diagonals---which characterize the between-class accessibility and will favor amenity class pairs that are highly mixed. As such, a low value of this average would simply indicate a high density of amenities, regardless of their class. (We will discuss further density-related considerations in Sec.~\ref{sec:results}.)    

With this in mind, we can construct a measure of the overall circuity of accessibility among amenities in a city by looking at an average over individual amenities rather than full amenity classes (to ensure appropriate weighting by class frequency), and only considering distinct amenity classes (off-diagonals of the PC matrix). The resulting measure, which we call the \emph{Average Circuity} (AC), can be written as
\begin{align}
\text{AC} &= \frac{1}{N}\sum_{i=1}^{N}\Big(\frac{1}{C-1}\sum_{\beta \neq c_i}\Big[d_G(i,j^\ast(i,\beta))\\
&~~~~~~~~~~~~~~~~~~~~~~~-d_E(i,j^\ast(i,\beta))\Big]\Big), \nonumber   
\end{align}
which can be interpreted as the expected excess travel distance from a randomly chosen amenity to the nearest amenity of a randomly chosen (different) class. This can be equivalently written in terms of the Pairwise Circuity matrix as
\begin{align}\label{eq:Average Circuity}
\text{AC} = \frac{1}{C-1}\sum_{\alpha=1}^{C}\frac{n_\alpha}{N}\sum_{\beta\neq\alpha }& \text{PC}_{\alpha\beta}.    
\end{align}
It is also useful to decompose the Average Circuity into the contributions from each individual amenity class $\alpha$, giving us the \emph{Marginal Circuity} (MC)

\begin{align}\label{eq:Marginal Circuity}
\text{MC}_\alpha = \frac{1}{C-1}\sum_{\beta\neq\alpha } & \text{PC}_{\alpha\beta},     
\end{align}
which evidently satisfies 
\begin{align}
\sum_{\alpha}\frac{n_\alpha}{N}\text{MC}_\alpha=\text{AC}.    
\end{align}

These aggregate measures, as well as the original Pairwise Circuity measure, are summarized in Table~\ref{tab-definitions}.

\section{Results}\label{sec:results}

\subsection{Correlations with Prosperity and Accessibility Indicators}\label{sec:correlations}

To examine the extent to which the circuity component of accessibility (i.e. the directness of accessibility) among urban amenities is associated with various facets of socioeconomic and environmental prosperity in cities, we compute the Spearman rank correlation between all pairs of Marginal Circuity values (Eq.~\ref{eq:Marginal Circuity}) and prosperity indicators discussed in Sec.~\ref{sec:data}. Since lower values of the MC correspond to more direct accessibility (less circuity), we expect positive correlations between our MC measures and development indicators where low values indicate high prosperity (e.g. the Gini coefficient and the poverty rate), and negative correlations with development indicators where high values indicate high prosperity (the other eight measures in Table S1, excluding WalkScore and BikeScore). In Fig.~\ref{Correlation} we show the results of these experiments in the first ten rows, and in Figures S4 and S5 we show scatterplots of the city indicators versus the Marginal Circuity values for the Health and Education amenity classes.

For each experiment, to ensure that amenity density and class frequency are not the primary factors determining the correlation, we compare the true Spearman rank correlation value with 100 randomized realizations of the city where all other factors are fixed but the amenity class labels of the amenities are shuffled at random while preserving their frequencies (see Sec.~\ref{sec:Pairwise Circuity} for details). These random realizations have the same frequency and average spatial density across the city for each amenity class, but the spatial and topological correlations among the amenities is destroyed. We then compute an empirical permutation test p-value representing the fraction of random realizations with Spearman rank correlations higher than the true observed correlation. Asterisks indicate (Marginal Circuity, Urban Indicator) pairs for which the empirical p-value was less than $0.05$, indicating that fewer than $5\%$ of the randomized trials produced correlations higher than the observed correlation. 

\begin{figure*}[t]
\includegraphics[width=16cm]{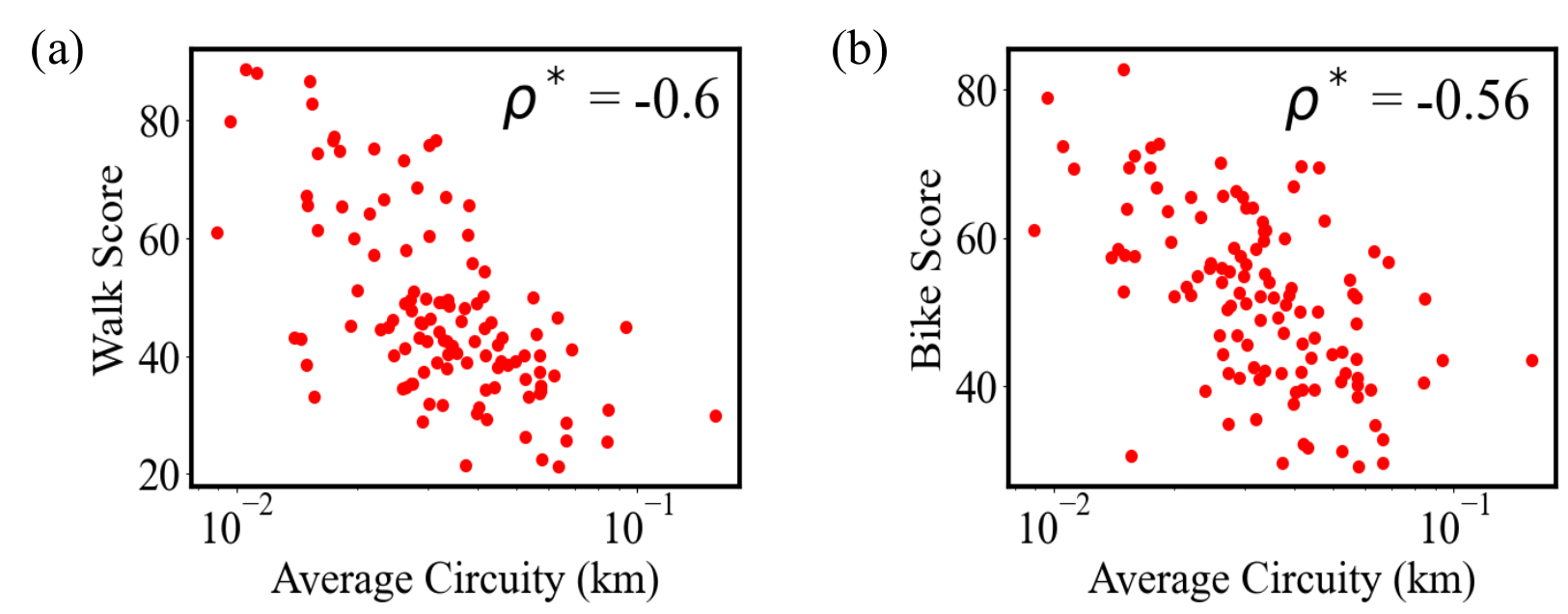}
    \caption{\textbf{Relationship between Average Circuity and Accessibility Indicators.} \textbf{(a)-(b)} Walk Score and Bike Score \cite{WalkScore} versus Average Circuity for 115 cities (Sec.~\ref{sec:data}). Superscript $*$ indicates a permutation test $p$-value of less than $0.05$ over 100 draws from the null model where the class labels of the amenities are shuffled while maintaining the overall class frequencies.}
    \label{Walk}
\end{figure*}

We observe a striking overall trend in the results suggesting that the circuity of accessibility from most amenity types is significantly associated with many of the prosperity indicators. As expected, we find a positive correlation when comparing the Gini coefficient and poverty rate with the Marginal Circuity of a number of amenities, indicating that cities whose street infrastructure does not facilitate efficient access among different amenities tend to have more income inequality and higher poverty rates. Meanwhile---also as expected---we find a negative correlation with the other prosperity metrics (e.g. quality of life and GDP per capita), indicating that as amenities become less directly accessible along the street infrastructure these indicators of prosperity and livability decline. 

We also repeat these experiments with the Walk Score and Bike Score indices (the last two rows in Fig.~\ref{Correlation}), to examine whether the circuity component of accessibility as measured by the Pairwise Circuity values is associated with the overall walkability and bikeability of a city according to these measures. Once again we observe negative correlations with all Marginal Circuity values, indicating that lower excess travel costs are associated with higher levels of walkability and bikeability in the studied cities. Taken together, the correlations identified in Fig.~\ref{Correlation} are consistent with the expectation that greater excess travel costs between amenities will be associated with lower levels of prosperity and accessibility.   

It is important to note that the definitions of the Walk Score and Bike Score \cite{walkscoremethodology,hall2018walk} do not necessitate a correlation with our MC measures, despite both being computed using the street network and various amenities. The Walk Score and Bike Score compute accessibility based on pure network distances, while our measure focuses on the difference between the network distance and the crow-fly distance. Due to high correlations with crow-fly distances, pure street network distances are primarily reflective of urban density rather than network efficiency/circuity, so it is not necessarily the case that cities with high Walk Scores and Bike Scores will have low scores according to our PC measures. A counterexample demonstrating this would be a city with a high density but a very inefficient, circuitous street network. Such a city will have a high Walk Score and Bike Score due to short distances between amenities along the street network. However, it may have a high PC score relative to a randomized null model where amenity labels are shuffled. This is because, despite being short in absolute terms, the network travel distance among amenities is much longer than expected based on their crow-fly distances. Therefore, our measure will highlight this network inefficiency, while the Walk Score and Bike Score will only highlight the overall amenity density. 

To check whether these effects are present when considering the average circuity of accessibility among all amenity classes in each city, in Fig.~\ref{Walk} and Fig.~S6 we examine the associations between the indicators and the Average Circuity values, finding that the trends mirror those seen for the Marginal Circuity. Thus, whether one marginalizes over specific amenities or considers them as a whole, the association between the information contained in the Pairwise Circuity matrix and city indicators is robust.

\subsection{Pairwise Circuity Distributions for Groups of Global Cities}

\begin{figure*}[t!]
\includegraphics[width=18cm]{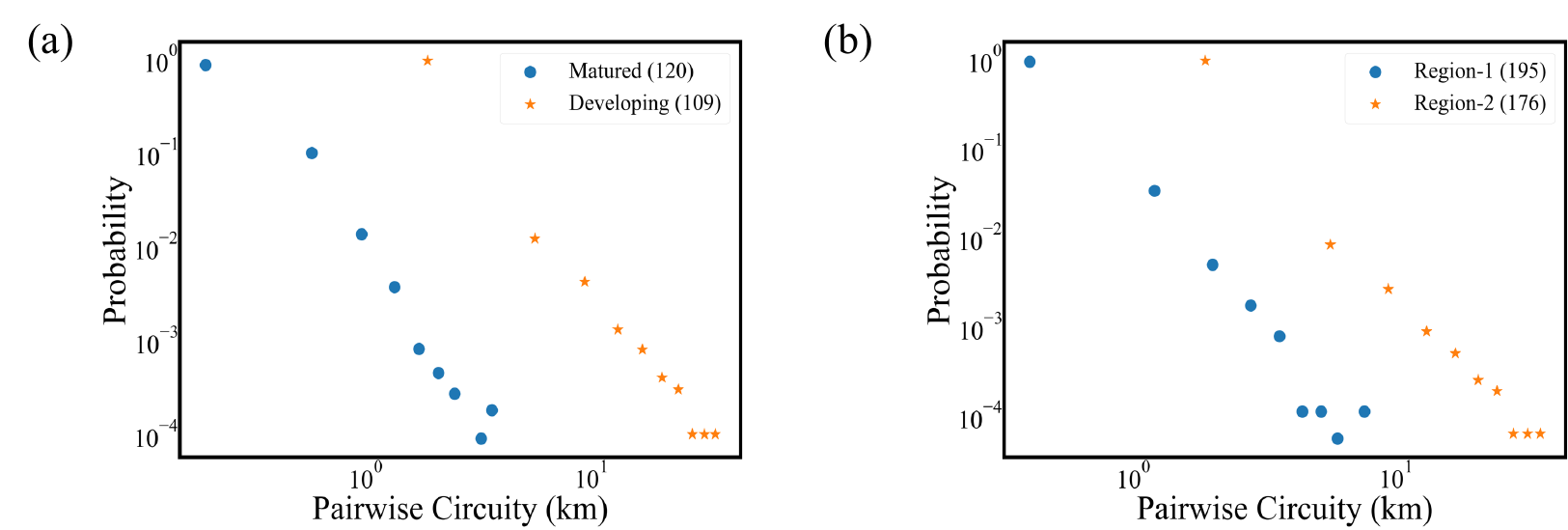}
    \caption{\textbf{Differences in Pairwise Circuity values across groups of global cities.} \textbf{(a)} The probability density of Pairwise Circuity (Eq.~\ref{Pairwise Circuity}) for cities in different development subgroups. \textbf{(b)} The probability density of Pairwise Circuity for cities in different regional subgroups (Sec.~\ref{sec:data}). The observed distributional orderings are statistically significant at the $p=0.05$ level when compared to the null model where class labels of the amenities are shuffled while maintaining the overall class frequencies. The numbers in the parentheses represent the number of cities in each city subgroup. The regions in panel (b) roughly correspond to the Global North (Region 1) and Global South (Region 2), as well as the division between developed (Region 1) and developing (Region 2) countries according to International Monetary Fund \cite{IMF}.}
    \label{Group}
\end{figure*}

Next we disaggregate the results to understand how the Pairwise Circuity varies across groups of cities at different stages of economic development---that is, Mature or Developing, or Region 1 and Region 2 corresponding to the Global North/South divide (see Sec.~\ref{sec:data}). In Fig.~\ref{Group} we plot the probability distributions for the Pairwise Circuity according to the two development levels (panel a) and two world regions (panel b). We observe a clear separation in the distributions of the Pairwise Circuity values in each case. For the development levels, we find that the Matured cities have systematically lower Pairwise Circuity values than the Developing cities, and correspondingly amenities in Region 1 are more directly accessible than those in Region 2, despite the actual distribution of amenities being the same regardless of classification. 

In order to determine the statistical significance of the observed discrepancies between the Pairwise Circuity distributions within each city subgroup in Fig.~\ref{Group}, we compute the Jonckheere trend test statistic \cite{jonckheere1954distribution} assuming an a priori ordering of the distributions---Matured $<$ Developing and Region 1 $<$ Region 2 in panels (a) and (b) respectively. The Jonckheere statistic $S$ measures the extent to which the values in the group hypothesized to have higher values ($d_{high}$) exceed the values in the group assumed to have lower values ($d_{low}$), and is given by
\begin{align}
S = \sum_{x\in d_{low}}\sum_{y\in d_{high}} [\bm{1}(x\leq y) - \bm{1}(x > y)],    
\end{align}
where $\bm{1}(\cdot)$ is the indicator function. (This is equivalent to a rescaled Mann-Whitney test statistic for two groups and no ties between the groups, although the Jonckheere statistic is applicable to cases with more than two ordered groups.) 

In its standard form, the Jonckheere trend test is performed by computing the p-value associated with the $S$ statistic under its corresponding standard normal approximation, in which case it has more statistical power than the Kruskal Wallis test for a priori ordered distributions \cite{jonckheere1954distribution}. Under this test, our results are highly statistically significant ($p\ll 0.01$ for the $S$ statistics of both panels (a) and (b)). However, we also perform an additional, more stringent test of statistical significance to ensure we can account for amenity density- and frequency-related effects. We compute the empirical p-value for the $S$ statistic associated with the real data by comparing this value with the $S$ statistics obtained for $100$ null model simulations. Specifically, in each simulation we shuffle the amenity labels within each city and calculate its resulting set of Pairwise Circuity values, then aggregate all these Pairwise Circuity values for the two city subgroups of interest. We then recompute the $S$ statistic for these two distributions of Pairwise Circuity values, and repeat the simulation process for $100$ trials. For both the development level comparison (panel a) and the world region comparison (panel b), we find empirical p-values of less than $p=0.05$, indicating that in more than $95\%$ of trials the null model resulted in a lower $S$ statistic than that which was observed for the real data. We note that this provides rather strong evidence of the Pairwise Circuity being a robust indicator of city prosperity, given that it tracks with the classification of cities done by external agencies (United Nations and JLL) using methods quite different than considered here \cite{UnHabitat,JLL}.

\section{Discussion}

In this paper we develop a simple, principled, and flexible framework to characterize the circuity of accessibility among amenities in a city based on the typical excess travel distance incurred to route between amenities of different types. Our method, which is built on what we call the Pairwise Circuity, simultaneously accounts for the density, heterogeneity, and adjacency of amenities along an underlying urban street network, effectively integrating both spatial and topological correlations among these points of interest in a single measure. We observe strong correlations between our Pairwise Circuity-based measures and various urban prosperity and accessibility indicators, as well as with the development levels and world regions of global cities. All of our results are robust when compared to a null model that scrambles the correlations among amenities while preserving their densities and frequencies throughout the city, confirming that the Pairwise Circuity measures provide a more nuanced view of accessibility beyond the density and diversity of amenities, which are the primary factors of interest in existing accessibility measures. 

A comprehensive analysis of the accessibility and vitality of cities requires the consideration of many distinct competing factors \cite{geurs2015accessibility,sung2015residential} and spatial scales \cite{talen2022can,fang2021spatial}, underscoring the importance of developing interpretable measures that can parsimoniously describe multifaceted structural correlations in cities that are relevant for accessibility and vitality. The results in our experiments suggest that the Pairwise Circuity framework can provide a complementary but distinct view of urban structure to existing measures of accessibility and urban vitality by assessing the circuity of accessibility among amenities along the street infrastructure conditioned on their spatial locations. Our measure is easy to compute for large-scale urban datasets and, when used in conjunction with the null model we present, provides a framework for assessing correlations among amenities that is robust to variations in their sampling density, allowing it to be used in data-scarce environments. Accessibility and diversity of amenities are key components of urban vitality \cite{jacobs2016death}, and circuity is a key component of accessibility along an infrastructure network \cite{huang2015circuity}, so by combining notions of circuity and diversity among urban amenities, the Pairwise Circuity measure we present captures a fundamental contributor to urban vitality that is often overlooked in favor of measures based on amenity diversity or density alone.

There are a number of ways in which the pairwise circuity framework developed in this paper can guide policy development aimed at improving urban street networks and amenities. For applications specifically concerned with developing street infrastructure that provides direct accessibility to/from a certain amenity type---e.g. healthcare amenities, whose accessibility is critical for rapidly aging urban societies---planners and policymakers can compute the pairwise circuity (PC) and/or marginal circuity (MC) measures for this focus amenity in different administrative regions. Areas with low PC/MC scores can then be targeted for interventions such as the construction of additional access roads to major highways or public transportation. On the other hand, if a practitioner aims to identify target areas for supplementing street infrastructure in a city more generally, they can use the whole PC matrix or the aggregated average circuity (AC) measure we propose to identify target regions. Finally, for global policy, large-scale analyses comparing the aggregate PC properties of cities (e.g. with the MC and AC measures we present) may be appropriate, as this can identify cities that are in need of better street infrastructure to facilitate direct accessibility among their urban amenities of various types.

Our measures can be extended in a number of meaningful ways in future work. In this work, given an origin amenity, we have identified the destination amenity within each class as the amenity minimizing the Euclidean distance to the starting point. However, in principle one can use any number of travel cost measures (e.g. travel time) or desirability measures (e.g. popularity) to identify the destination amenity within each class. One can also use a non-deterministic mechanism such as a stochastic route choice model or biased random walk to choose the destination amenities to which we compute the excess travel costs for each origin. One could additionally perform the Pairwise Circuity analysis using a more generalized travel cost measure for both the crow-fly and street network paths, resulting in a measure with units of cost rather than distance.\\

\section*{Author Contributions}
B.P.: Conceptualization, data curation, computational analysis, writing.
G.G.: Conceptualization, supervision, writing.
A.K.: Conceptualization, methodology, supervision, writing.  

\section*{Acknowledgements}
The authors would like to thank Ronaldo Menezes for useful discussions. 

\section*{Data Accessibility}
All data are publicly available online from the cited sources, as well as available from the authors upon request. 

\section*{Funding Statement} 
A.K. was funded in part by the HKU-100 Start Up Fund. B.P. and G.G. thank the University of Rochester for support. 

\providecommand{\noopsort}[1]{}\providecommand{\singleletter}[1]{#1}%


\clearpage
\onecolumngrid

\setcounter{equation}{0}
\setcounter{figure}{0}
\setcounter{section}{0}
\setcounter{table}{0}
\setcounter{page}{1}
\renewcommand{\theequation}{S\arabic{equation}}
\renewcommand{\thefigure}{S\arabic{figure}}
\renewcommand{\thesection}{\arabic{section}}
\renewcommand{\thetable}{S\arabic{table}}
\renewcommand{\bibnumfmt}[1]{[S#1]}
\renewcommand{\citenumfont}[1]{S#1}

\begin{center}
    \large{\textbf{Supplementary Material}}
\end{center}

\section{Cities with socioeconomic information}
    
Figures S1 and S2 display the 371 cities studied in our analyses, which collectively contain around 80\% of the  headquarters of the 2,000 largest companies in world, account for roughly 1.5 billion people, and 40\% of economic activity worldwide.
JLL source: \url{https://seoulsolution.kr/sites/default/files/gettoknowus/jll-global300-2015.pdf}, UN source: \url{https://data.unhabitat.org/pages/datasets} and WalkScore source: \url{https://www.walkscore.com/cities-and-neighborhoods}.
\begin{figure}[H]
\centering
    \includegraphics[width=12cm]{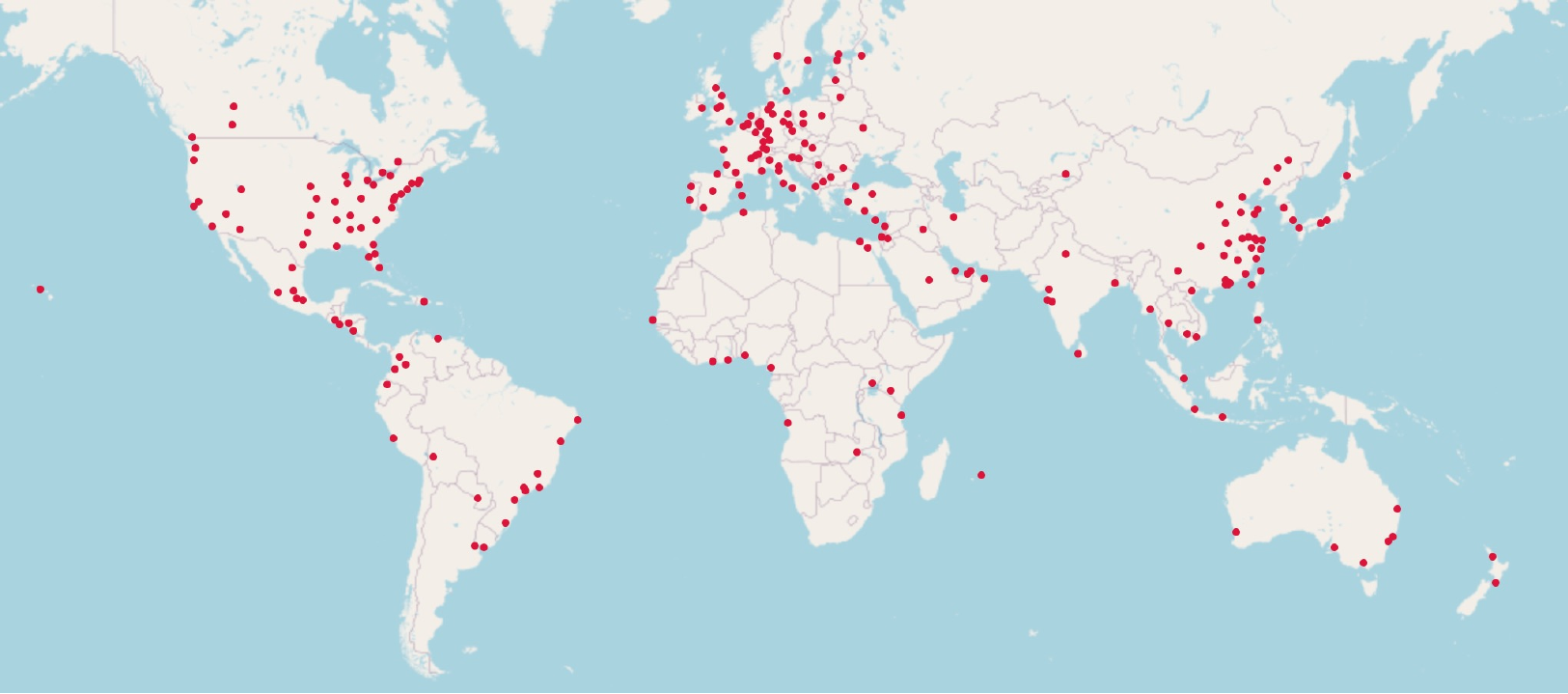}
    \caption{Map of cities used in the study.}
\end{figure}

\begin{figure}[H]
\centering
    \includegraphics[width=18cm]{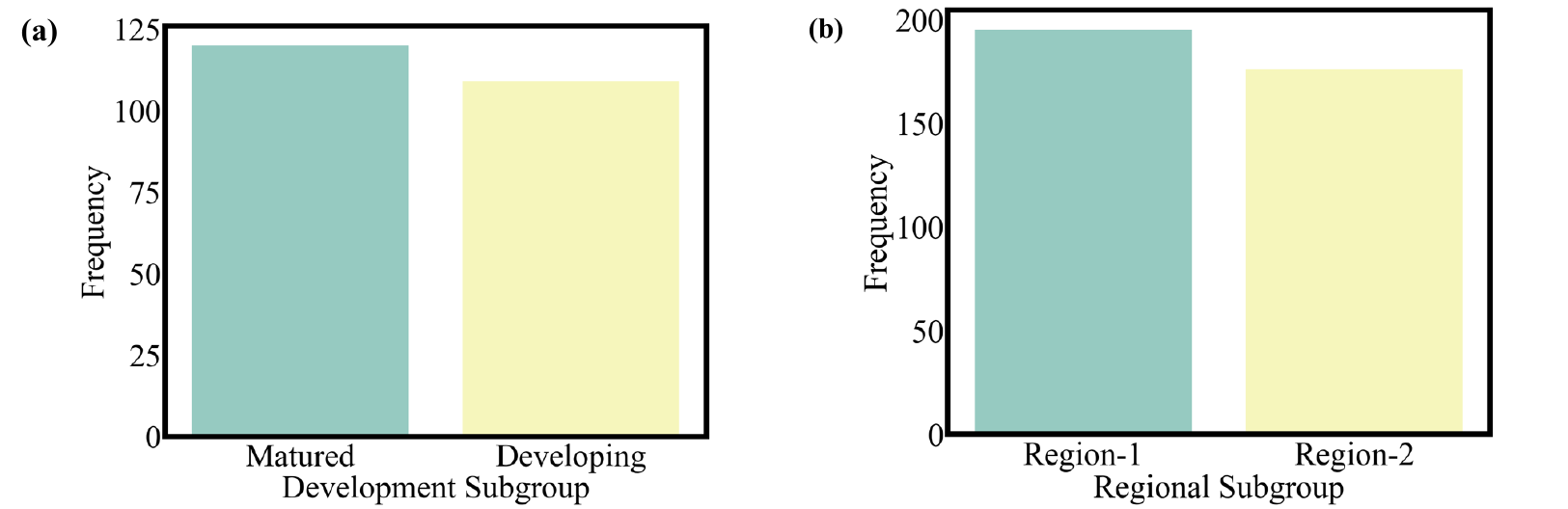}
    \caption{Number of cities in each development stage and world region.}
\end{figure}

\clearpage
\section{City Indicators}
In this paper we studied 10 prosperity indicators and 2 accessibility indicators. We collected prosperity indicator data from several UN sources and accessibility indicator data from the $\text{Walk Score}^{\textregistered}$ website, listed above. Table 1 below provides a description of each indicator studied.
\begin{table}[H]
\begin{tabular}{ | p{3cm}| p{12cm}| p{2cm}|}
 \hline
 \multicolumn{3}{|c|}{\textbf{City Indicators}} \\
 \hline
\textbf{Indicator} & \textbf{Definition} & \textbf{\# of Cities}\\
 \hline
Gini Coefficient & Standard measure of economic disparity, ranging from $0$ (perfect equality) to $1$ (complete inequality). & 49 
\\
\hline
Internet Access & Percentage of the population with access to the internet. & 72 
 \\
 \hline
 Public Transportation Access & Percentage of population with access to public transportation. & 153 
 \\
 \hline
 GDP Per Capita & Total gross domestic product for a city divided by population. & 49
 \\
 \hline
QL Index & Quality of Life Index, an aggregate measure of quality of health, education and security. & 49
 \\
 \hline
Poverty Rate & Percentage of the population whose income is below the poverty line. & 71
 \\
 \hline
Infrastructure Index & Overall measurement of quality of housing infrastructure, social infrastructure, internet access, and urban mobility. & 49
 \\
 \hline
LOSI & Local Online Service Index, a measure ranging from $0$ to $1$ that assesses the services and information provided by local governments through their websites. & 135
\\
\hline
Public Space & Percentage of land allocated for open space. & 72  
 \\
\hline
Public Space Access & Percentage of the population with access to land allocated for open space. & 72 
 \\
\hline
 Walk Score & Measure of walkability that analyzes several viable walking routes to different amenities. & 115
 \\
 \hline
 Bike Score & Similar to Walk Score, but for bikeability. & 115
 \\
 \hline
\end{tabular}
\caption{Definitions of the city indicators used in this paper, along with the number of cities analyzed for each indicator.} 
\label{tab-indicators}
\end{table}
\clearpage

\section{Distribution of city diameter, average shortest path length, and amenity density across city subgroups}
 \begin{center}
 \begin{figure}[H]
 \includegraphics[width=18cm]{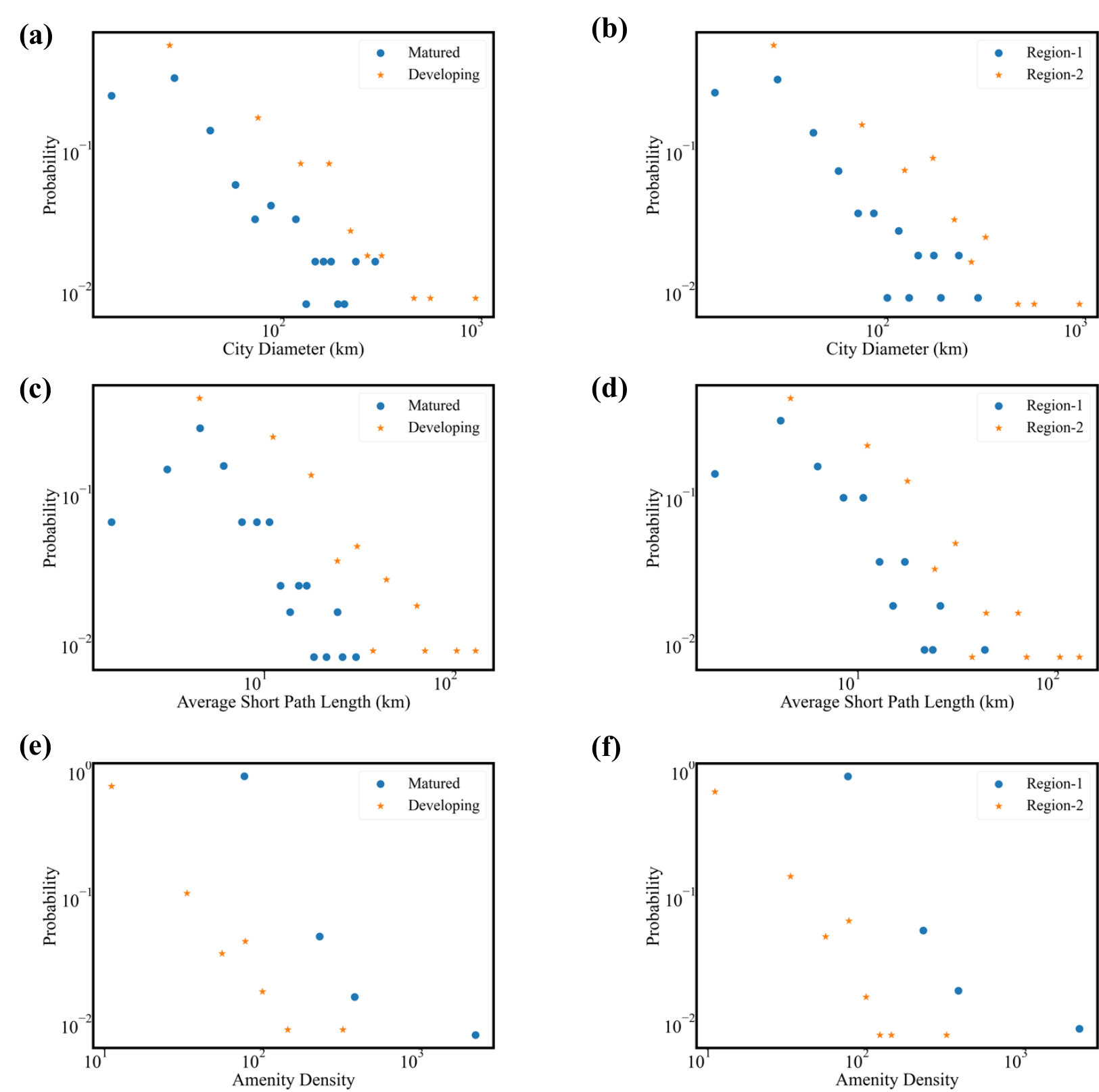}
    \caption{Distributions of city diameter (top), average short path length (middle) and amenity density (bottom), for the city groupings used in this study.}
\end{figure}
\end{center}

\clearpage
\section{Scatterplots of Marginal Circuity $\text{MC}_{Hlt.}$ versus city indicators}

\begin{figure}[H]
\centering
 \includegraphics[width=18cm]{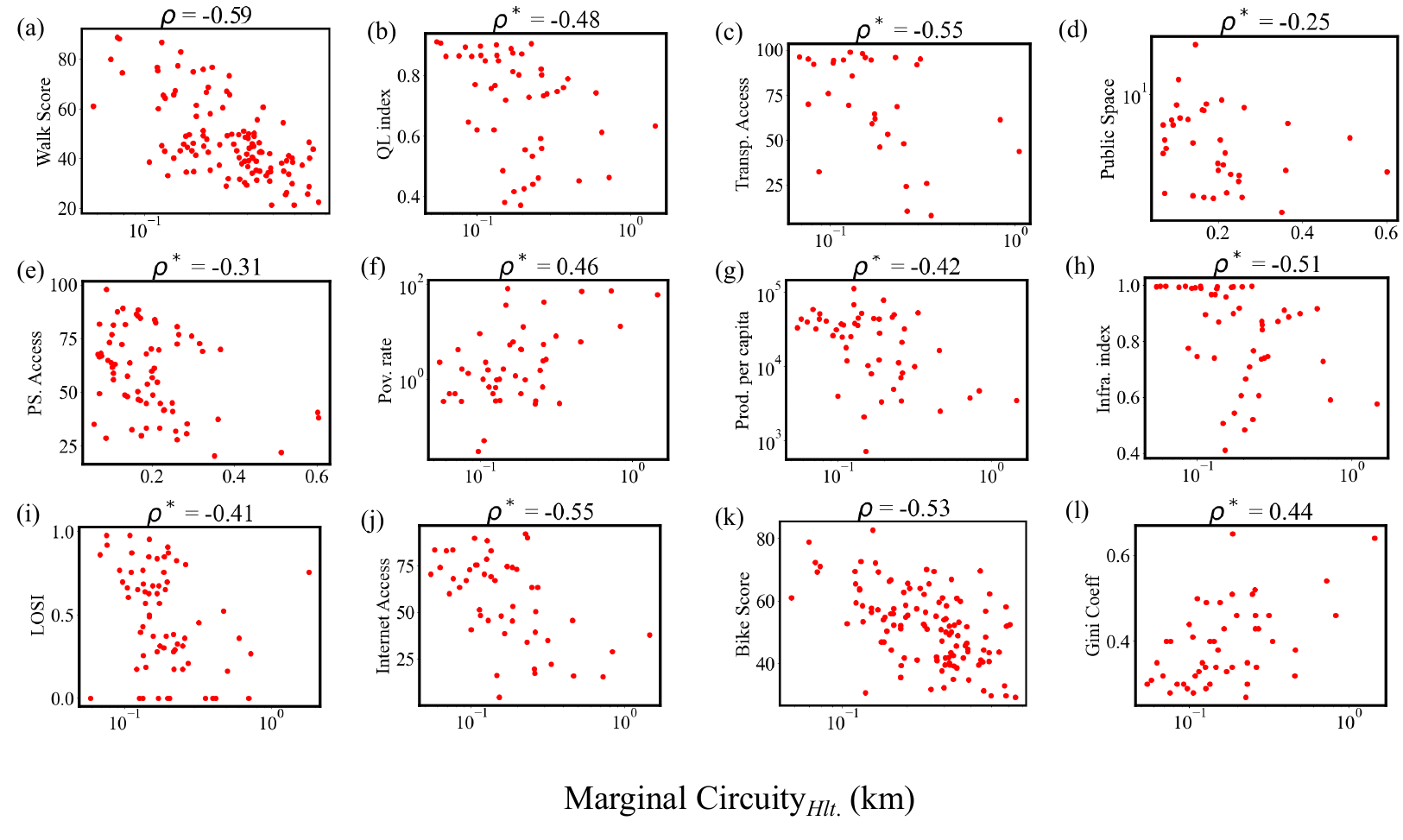}
    \caption{Health Marginal Circuity versus prosperity and accessibility indicators. Superscript $*$ indicates a permutation test $p$-value of less than $0.05$ over 100 draws from the null model where the class labels of the amenities are shuffled while maintaining the overall class frequencies.}
    \label{Fin}
\end{figure}

\clearpage
\section{Scatterplots of Marginal Circuity $\text{MC}_{Ed.}$ versus city indicators}

\begin{figure}[H]

\centering
\includegraphics[width=18cm]{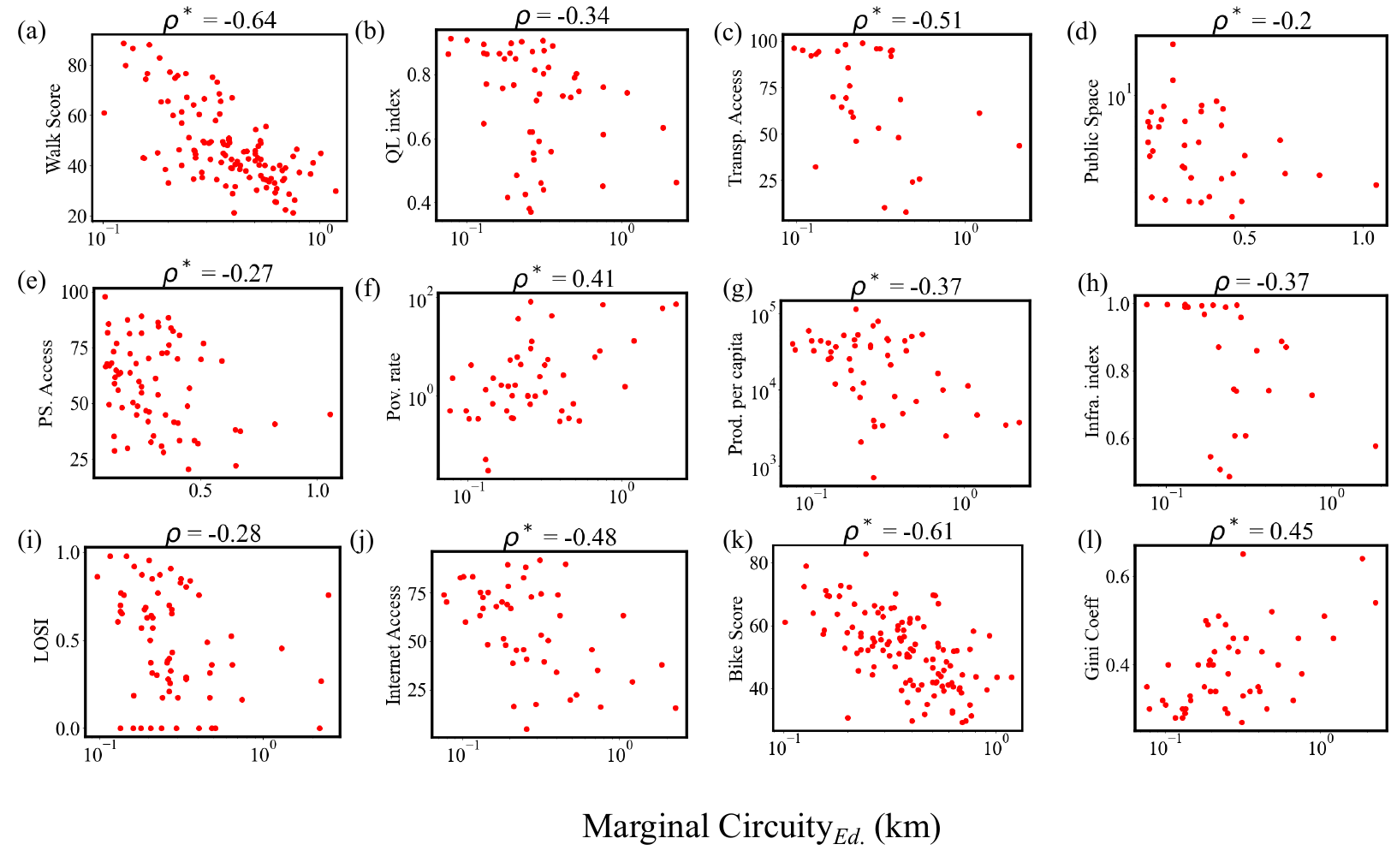}
    \caption{Education Marginal Cicuity versus prosperity and accessibility indicators. Superscript $*$ indicates a permutation test $p$-value of less than $0.05$ over 100 draws from the null model where the class labels of the amenities are shuffled while maintaining the overall class frequencies.}
    \label{Ed}
\end{figure}

\newpage
\section{Average Circuity}

\begin{figure}[H]
\centering
\includegraphics[width=18cm]{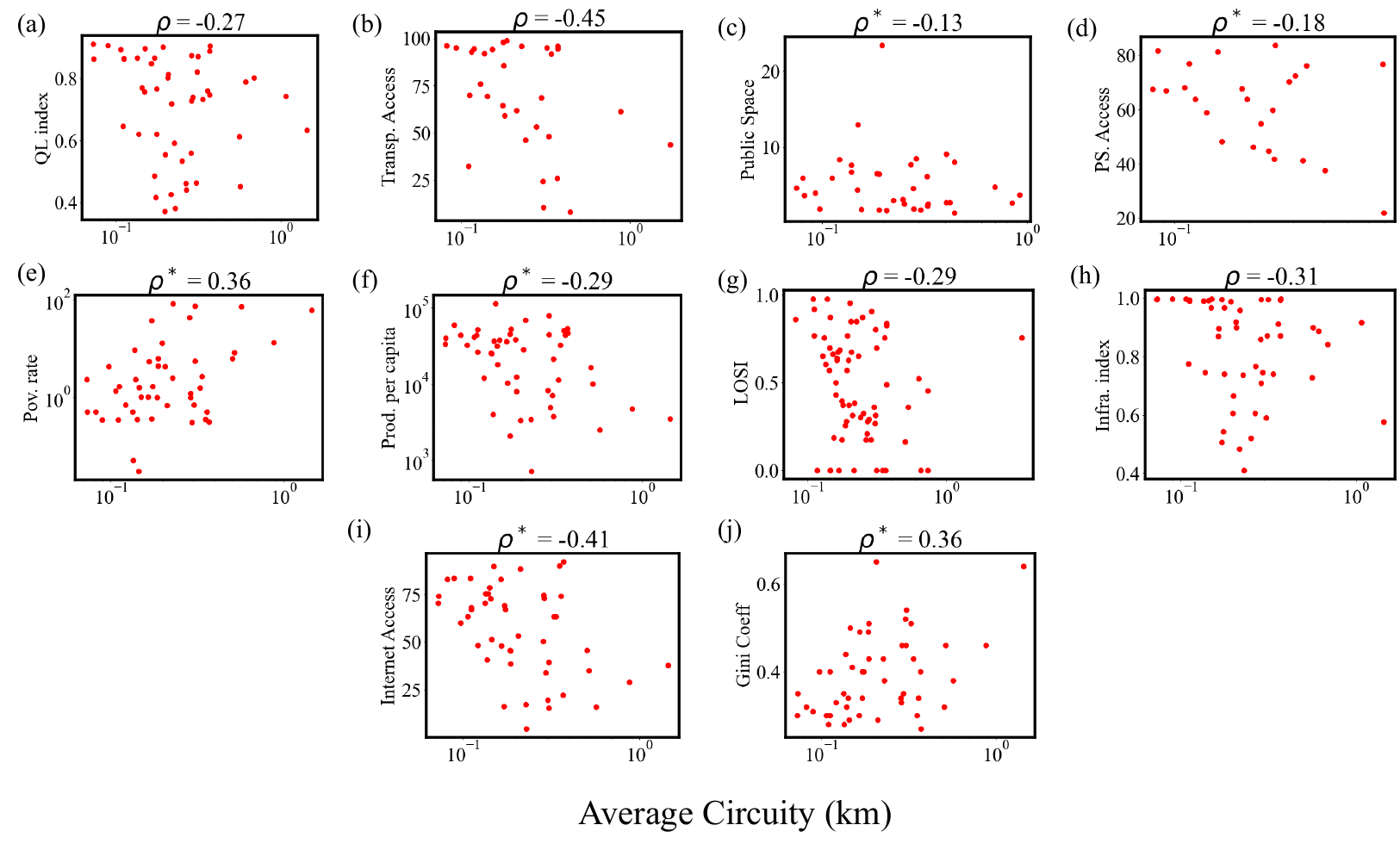}
    \caption{Relationship of Average Circuity with prosperity indicators. Superscript $*$ indicates a permutation test $p$-value of less than $0.05$ over 100 draws from the null model where the class labels of the amenities are shuffled while maintaining the overall class frequencies.}
    \label{Average Circuity}
\end{figure}

\clearpage
\section{Difference in NYC Pairwise Circuity matrices on a diverging color scale}
\begin{figure}[H]
\centering
\includegraphics[width=10cm]{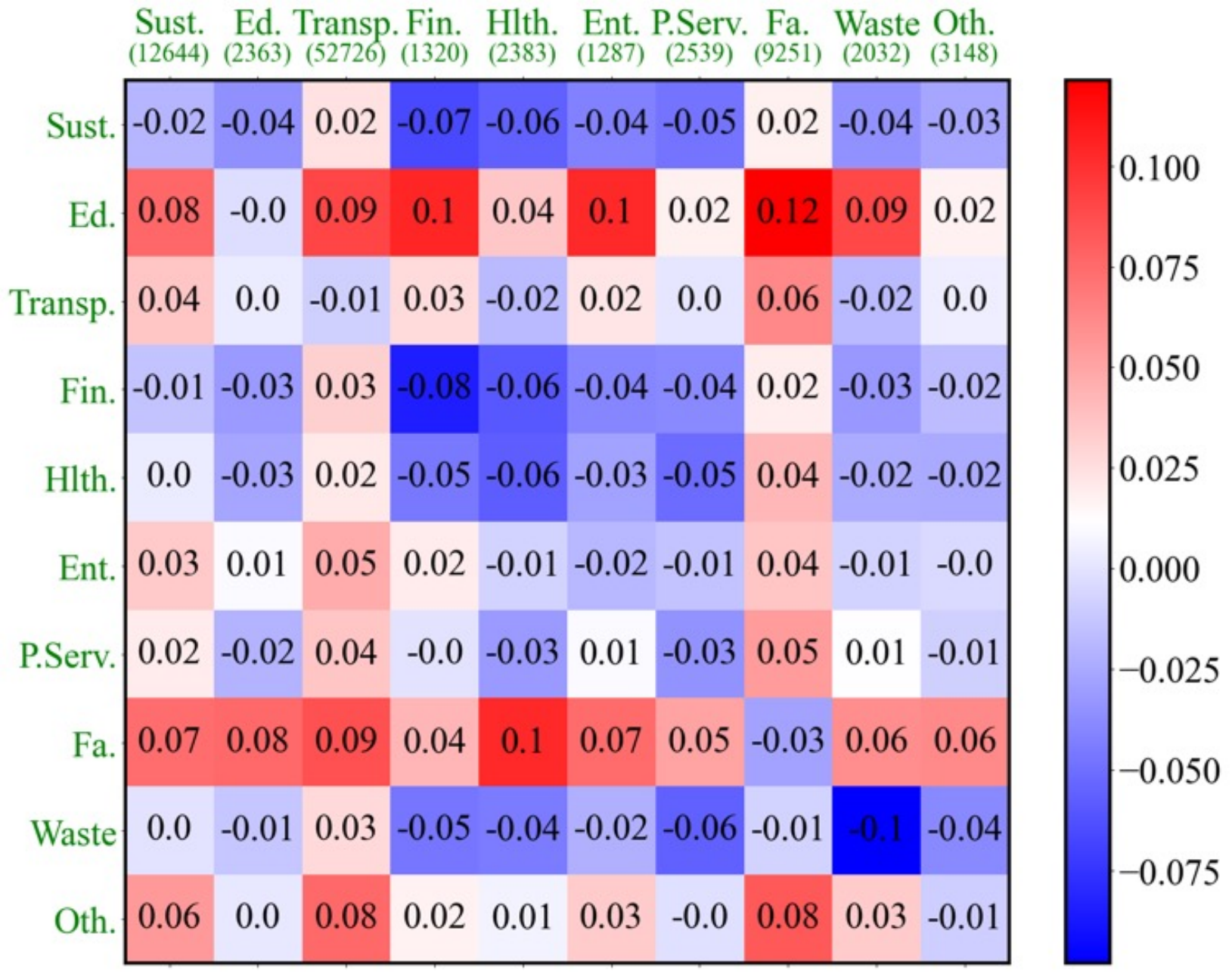}
    \caption{The differences in Pairwise Circuity obtained from observed amenity labels and null model simulations in NYC, on a diverging color scale. All values are in the units of km. In this case, shortest paths originating at the amenities in the classes `Education', `Facilities', and `Other' tend to be more circuitous (higher PC) than expected by random chance, and all others tend to be less circuitous (lower PC) than expected in the null model.}
    \label{RR3}
\end{figure}


\begin{thebibliography}{10}
\expandafter\ifx\csname url\endcsname\relax
  \def\url#1{\texttt{#1}}\fi
\expandafter\ifx\csname urlprefix\endcsname\relax\def\urlprefix{URL }\fi

\bibitem{barthelemy2016structure}
M.~Barthelemy, \textit{The Structure and Dynamics of Cities}. Cambridge
  University Press (2016).

\bibitem{bettencourt21}
L.~M.~A. Bettencourt, \textit{Introduction to Urban Science: Evidence and
  Theory of Cities as Complex Systems}. MIT Press, Cambridge, MA (2021).

\bibitem{boeing2018measuring}
G.~Boeing, Measuring the Complexity of Urban Form and Design. \textit{Urban
  Design International} \textbf{23}(4), 281--292 (2018).

\bibitem{Pan2013}
W.~Pan, G.~Ghoshal, C.~Krumme, M.~Cebrian, and A.~Pentland, {Urban
  characteristics attributable to density-driven tie formation}. \textit{Nature
  Communications} \textbf{4}, 1--7 (2013).

\bibitem{guite2006impact}
H.~F. Guite, C.~Clark, and G.~Ackrill, The Impact of the Physical and Urban
  Environment on Mental Well-being. \textit{Public Health} \textbf{120}(12),
  1117--1126 (2006).

\bibitem{witten2003quality}
K.~Witten, D.~Exeter, and A.~Field, The Quality of Urban Environments: Mapping
  Variation in Access to Community Resources. \textit{Urban studies}
  \textbf{40}(1), 161--177 (2003).

\bibitem{king2015disadvantaged}
K.~E. King and P.~J. Clarke, A Disadvantaged Advantage in Walkability: Findings
  from Socioeconomic and Geographical Analysis of National Built Environment
  Data in the United States. \textit{American Journal of Epidemiology}
  \textbf{181}(1), 17--25 (2015).

\bibitem{bereitschaft2017equity}
B.~Bereitschaft, Equity in Neighbourhood Walkability? A Comparative Analysis of
  Three Large US Cities. \textit{Local Environment} \textbf{22}(7), 859--879
  (2017).

\bibitem{Bassolas2019}
A.~Bassolas, H.~Barbosa-Filho, B.~Dickinson, X.~Dotiwalla, P.~Eastham,
  R.~Gallotti, G.~Ghoshal, B.~Gipson, S.~A. Hazarie, H.~Kautz, O.~Kucuktunc,
  A.~Lieber, A.~Sadilek, and J.~J. Ramasco, {Hierarchical organization of urban
  mobility and its connection with city livability}. \textit{Nature
  Communications} \textbf{10}(1), 1--10 (2019),
  \urlprefix\url{http://dx.doi.org/10.1038/s41467-019-12809-y}.

\bibitem{kandt2021smart}
J.~Kandt and M.~Batty, Smart Cities, Big Data and Urban Policy: Towards Urban
  Analytics for the Long Run. \textit{Cities} \textbf{109}, 102992 (2021).

\bibitem{boeing2020multi}
G.~Boeing, A Multi-scale Analysis of 27,000 Urban Street Networks: Every US
  City, Town, Urbanized Area, and Zillow Neighborhood. \textit{Environment and
  Planning B: Urban Analytics and City Science} \textbf{47}(4), 590--608
  (2020).

\bibitem{kitchin2015knowing}
R.~Kitchin, T.~P. Lauriault, and G.~McArdle, Knowing and Governing Cities
  through Urban Indicators, City Benchmarking and Real-time Dashboards.
  \textit{Regional Studies, Regional Science} \textbf{2}(1), 6--28 (2015).

\bibitem{jacobs2016death}
J.~Jacobs, \textit{The Death and Life of Great American Cities}. Vintage
  (2016).

\bibitem{eom2020land}
S.~Eom, T.~Suzuki, and M.-H. Lee, A Land-use Mix Allocation Model Considering
  Adjacency, Intensity, and Proximity. \textit{International Journal of
  Geographical Information Science} \textbf{34}(5), 899--923 (2020).

\bibitem{lu2019impacts}
S.~Lu, C.~Shi, and X.~Yang, Impacts of Built Environment on Urban Vitality:
  Regression Analyses of Beijing and Chengdu, China. \textit{International
  Journal of Environmental Research and Public health} \textbf{16}(23), 4592
  (2019).

\bibitem{sung2015residential}
H.~Sung and S.~Lee, Residential Built Environment and Walking Activity:
  Empirical Evidence of Jane Jacobs' Urban Vitality. \textit{Transportation
  Research Part D: Transport and Environment} \textbf{41}, 318--329 (2015).

\bibitem{yuan2012discovering}
J.~Yuan, Y.~Zheng, and X.~Xie, Discovering Regions of Different Functions in a
  City Using Human Mobility and POIs. In \textit{Proceedings of the 18th ACM
  SIGKDD International Conference on Knowledge Discovery and Data mining}, pp.
  186--194 (2012).

\bibitem{zhu2017building}
J.~Zhu and Y.~Sun, Building an Urban Spatial Structure from Urban Land Use
  Data: An Example Using Automated Recognition of the City Centre.
  \textit{ISPRS International Journal of Geo-Information} \textbf{6}(4), 122
  (2017).

\bibitem{lee2017morphology}
M.~Lee, H.~Barbosa, H.~Youn, P.~Holme, and G.~Ghoshal, Morphology of Travel
  Routes and the Organization of Cities. \textit{Nature Communications}
  \textbf{8}(1), 2229 (2017).

\bibitem{kirkley2022spatial}
A.~Kirkley, Spatial Regionalization Based on Optimal Information Compression.
  \textit{Communications Physics} \textbf{5}(1), 249 (2022).

\bibitem{kirkley2020information}
A.~Kirkley, Information Theoretic Network Approach to Socioeconomic
  Correlations. \textit{Physical Review Research} \textbf{2}(4), 043212 (2020).

\bibitem{arcaute2016cities}
E.~Arcaute, C.~Molinero, E.~Hatna, R.~Murcio, C.~Vargas-Ruiz, A.~P. Masucci,
  and M.~Batty, Cities and Regions in Britain through Hierarchical Percolation.
  \textit{Royal Society Open Science} \textbf{3}(4), 150691 (2016).

\bibitem{smoyer2004spatial}
K.~E. Smoyer-Tomic, J.~N. Hewko, and M.~J. Hodgson, Spatial Accessibility and
  Equity of Playgrounds in Edmonton, Canada. \textit{Canadian Geographer/Le
  G{\'e}ographe Canadien} \textbf{48}(3), 287--302 (2004).

\bibitem{das2021access}
A.~Das, M.~Das, and H.~Barman, Access to Basic Amenities and Services to Urban
  Households in West Bengal: Does its Location and Size of Settlements Matter?
  \textit{GeoJournal} \textbf{86}, 885--913 (2021).

\bibitem{burdziej2019using}
J.~Burdziej, Using Hexagonal Grids and Network Analysis for Spatial
  Accessibility Assessment in Urban Environments--a Case Study of Public
  Amenities in Toru{\'n}. \textit{Miscellanea Geographica} \textbf{23}(2),
  99--110 (2019).

\bibitem{smith2010neighbourhood}
D.~M. Smith, S.~Cummins, M.~Taylor, J.~Dawson, D.~Marshall, L.~Sparks, and
  A.~S. Anderson, Neighbourhood Food Environment and Area Deprivation: Spatial
  Accessibility to Grocery Stores Selling Fresh Fruit and Vegetables in Urban
  and Rural Settings. \textit{International Journal of Epidemiology}
  \textbf{39}(1), 277--284 (2010).

\bibitem{hewko2002measuring}
J.~Hewko, K.~E. Smoyer-Tomic, and M.~J. Hodgson, Measuring Neighbourhood
  Spatial Accessibility to Urban Amenities: Does Aggregation Error Matter?
  \textit{Environment and Planning A} \textbf{34}(7), 1185--1206 (2002).

\bibitem{tannier2012spatial}
C.~Tannier, G.~Vuidel, H.~Houot, and P.~Frankhauser, Spatial Accessibility to
  Amenities in Fractal and Nonfractal Urban Patterns. \textit{Environment and
  Planning B: Planning and Design} \textbf{39}(5), 801--819 (2012).

\bibitem{yuan2020amenity}
F.~Yuan, Y.~D. Wei, and J.~Wu, Amenity Effects of Urban Facilities on Housing
  Prices in China: Accessibility, Scarcity, and Urban Spaces. \textit{Cities}
  \textbf{96}, 102433 (2020).

\bibitem{logan2021measuring}
T.~M. Logan, M.~J. Anderson, T.~G. Williams, and L.~Conrow, Measuring
  Inequalities in Urban Systems: An Approach for Evaluating the Distribution of
  Amenities and Burdens. \textit{Computers, Environment and Urban Systems}
  \textbf{86}, 101590 (2021).

\bibitem{talen2022can}
E.~Talen, Who Can walk? An Analysis of Public Amenity Access in America's Ten
  Largest Cities. \textit{Environment and Planning B: Urban Analytics and City
  Science} p. 23998083221142866 (2022).

\bibitem{fang2021spatial}
C.~Fang, S.~He, and L.~Wang, Spatial Characterization of Urban Uitality and the
  Association with Various Street Network Metrics from the Multi-scalar
  Perspective. \textit{Frontiers in Public Health} \textbf{9}, 677910 (2021).

\bibitem{sulis2018using}
P.~Sulis, E.~Manley, C.~Zhong, and M.~Batty, Using Mobility Data as Proxy for
  Measuring Urban Vitality. \textit{Journal of Spatial Information Science}
  \textbf{16}, 137--162 (2018).

\bibitem{sulis2019measuring}
P.~Sulis, \textit{Measuring Urban Vitality through Human Mobility Patterns}.
  Ph.D. thesis, UCL (University College London) (2019).

\bibitem{kim2020urban}
S.~Kim, Urban Vitality, Urban Form, and Land Use: Their Relations within a
  Geographical Boundary for Walkers. \textit{Sustainability} \textbf{12}(24),
  10633 (2020).

\bibitem{zhang2021can}
A.~Zhang, W.~Li, J.~Wu, J.~Lin, J.~Chu, and C.~Xia, How Can the Urban Landscape
  Affect Urban Vitality at the Street Block Level? A Case Study of 15
  Metropolises in China. \textit{Environment and Planning B: Urban Analytics
  and City Science} \textbf{48}(5), 1245--1262 (2021).

\bibitem{xia2020analyzing}
C.~Xia, A.~G.-O. Yeh, and A.~Zhang, Analyzing Spatial Relationships Between
  Urban Land Use Intensity and Urban Vitality at Street Block Level: A Case
  Study of Five Chinese Megacities. \textit{Landscape and Urban Planning}
  \textbf{193}, 103669 (2020).

\bibitem{olszewski2005using}
P.~Olszewski and S.~S. Wibowo, Using Equivalent Walking Distance to Assess
  Pedestrian Accessibility to Transit Stations in Singapore.
  \textit{Transportation Research Record} \textbf{1927}(1), 38--45 (2005).

\bibitem{rastogi2003defining}
R.~Rastogi and K.~K. Rao, Defining Transit Accessibility with Environmental
  Inputs. \textit{Transportation Research Part D: Transport and Environment}
  \textbf{8}(5), 383--396 (2003).

\bibitem{geurs2015accessibility}
K.~Geurs and D.~Halden, Accessibility: Theory and Practice in the Netherlands
  and UK. In \textit{Handbook on Transport and Development}, pp. 459--476,
  Edward Elgar Publishing (2015).

\bibitem{landex2006examining}
A.~Landex and S.~Hansen, Examining the Potential Travellers in Catchment Areas
  for Public Transport. In \textit{ESRI User Conference} (2006).

\bibitem{garcia2018analysing}
J.~C. Garc{\'\i}a-Palomares, J.~S. Ribeiro, J.~Guti{\'e}rrez, T.~S. Marques,
  \textit{et~al.}, Analysing Proximity to Public Transport: the Role of Street
  Network Design. \textit{Bolet{\'\i}n de la Asociaci{\'o}n de Ge{\'o}grafos
  Espa{\~n}oles} (76), 102--130 (2018).

\bibitem{boscoe2012nationwide}
F.~P. Boscoe, K.~A. Henry, and M.~S. Zdeb, A Nationwide Comparison of Driving
  Distance versus Straight-line Distance to Hospitals. \textit{The Professional
  Geographer} \textbf{64}(2), 188--196 (2012).

\bibitem{bell201310}
N.~Bell, 10 Location--allocation Modelling for Health Services Research in Low
  Resource Settings. \textit{Geographic Health Data: Fundamental Techniques for
  Analysis} p. 165 (2013).

\bibitem{henry2013geographic}
K.~Henry, K.~McDonald, \textit{et~al.}, Geographic Access to Health Services.
  \textit{Geographic Health Data: Fundamental Techniques for Analysis; Boscoe,
  FP, Ed} pp. 142--164 (2013).

\bibitem{yenisetty2020spatial}
P.~T. Yenisetty and P.~Bahadure, Spatial Accessibility Measures to Educational
  facilities from public transit: a case of Indian cities. \textit{Smart and
  Sustainable Built Environment} \textbf{10}(2), 258--273 (2020).

\bibitem{he2019spatial}
S.~He, S.~Yu, P.~Wei, and C.~Fang, A Spatial Design Network Analysis of Street
  Networks and the Locations of Leisure Entertainment Activities: A Case Study
  of Wuhan, China. \textit{Sustainable Cities and Society} \textbf{44},
  880--887 (2019).

\bibitem{teeuwen2023measuring}
R.~Teeuwen, A.~Psyllidis, and A.~Bozzon, Measuring Children's and Adolescents'
  Accessibility to Greenspaces from Different Locations and Commuting Settings.
  \textit{Computers, Environment and Urban Systems} \textbf{100}, 101912
  (2023).

\bibitem{zhao2018towards}
Y.~Zhao, G.~Zhang, T.~Lin, X.~Liu, J.~Liu, M.~Lin, H.~Ye, and L.~Kong, Towards
  Sustainable Urban Communities: A Composite Spatial Accessibility Assessment
  for Residential Suitability Based on Network Big Data.
  \textit{Sustainability} \textbf{10}(12), 4767 (2018).

\bibitem{tu2020portraying}
W.~Tu, T.~Zhu, J.~Xia, Y.~Zhou, Y.~Lai, J.~Jiang, and Q.~Li, Portraying the
  Spatial Dynamics of Urban Vibrancy Using Multisource Urban Big Data.
  \textit{Computers, Environment and Urban Systems} \textbf{80}, 101428 (2020).

\bibitem{li2021multidimensional}
Q.~Li, C.~Cui, F.~Liu, Q.~Wu, Y.~Run, and Z.~Han, Multidimensional Urban
  Vitality on Streets: Spatial Patterns and Influence Factor Identification
  Using Multisource Urban data. \textit{ISPRS International Journal of
  Geo-Information} \textbf{11}(1), 2 (2021).

\bibitem{zhang2021analysis}
Y.~Zhang, L.~Yang, and X.~Wang, Analysis and Calculating of Comprehensive Urban
  Vitality Index by Multi-Source Temporal-Spatial Big Data and EW-TOPSIS. In
  \textit{2021 IEEE International Conference on Data Science and Computer
  Application (ICDSCA)}, pp. 196--201, IEEE (2021).

\bibitem{bassolas2021first}
A.~Bassolas and V.~Nicosia, First-passage Times to Quantify and Compare
  Structural Correlations and Heterogeneity in Complex Systems.
  \textit{Communications Physics} \textbf{4}(1), 76 (2021).

\bibitem{redner2001guide}
S.~Redner, \textit{A Guide to First-Passage Processes}. Cambridge University
  Press (2001).

\bibitem{bassolas2021diffusion}
A.~Bassolas, S.~Sousa, and V.~Nicosia, Diffusion Segregation and the
  Disproportionate Incidence of COVID-19 in African American Communities.
  \textit{Journal of the Royal Society Interface} \textbf{18}(174), 20200961
  (2021).

\bibitem{bongiorno2021vector}
C.~Bongiorno, Y.~Zhou, M.~Kryven, D.~Theurel, A.~Rizzo, P.~Santi, J.~Tenenbaum,
  and C.~Ratti, Vector-based Pedestrian Navigation in Cities. \textit{Nature
  Computational Science} \textbf{1}(10), 678--685 (2021).

\bibitem{miranda2021desirable}
A.~S. Miranda, Z.~Fan, F.~Duarte, and C.~Ratti, Desirable Streets: Using
  Deviations in Pedestrian Trajectories to Measure the Value of the Built
  Environment. \textit{Computers, Environment and Urban Systems} \textbf{86},
  101563 (2021).

\bibitem{barthelemy2011spatial}
M.~Barth{\'e}lemy, Spatial Networks. \textit{Physics Reports}
  \textbf{499}(1-3), 1--101 (2011).

\bibitem{boeing2017osmnx}
G.~Boeing, OSMnx: New methods for acquiring, constructing, analyzing, and
  visualizing complex street networks. \textit{Computers, Environment and Urban
  Systems} \textbf{65}, 126--139 (2017).

\bibitem{OpenStreetMap}
{OpenStreetMap contributors}, {Planet dump retrieved from
  https://planet.osm.org }. \url{ https://www.openstreetmap.org } (2017).

\bibitem{chen2021delineating}
Q.~Chen and A.~T. Crooks, Delineating a `15-minute city' an Agent-based
  Modeling Approach to Estimate the Size of Local Communities. In
  \textit{Proceedings of the 4th ACM SIGSPATIAL International Workshop on
  GeoSpatial Simulation}, pp. 29--37 (2021).

\bibitem{hu2016impacts}
N.~Hu, E.~F. Legara, K.~K. Lee, G.~G. Hung, and C.~Monterola, Impacts of Land
  Use and Amenities on Public Transport Use, Urban Planning and Design.
  \textit{Land Use Policy} \textbf{57}, 356--367 (2016).

\bibitem{s2022alf}
A.~M. S.~Alfosool, Y.~Chen, and D.~Fuller, ALF--Score---A novel Approach to
  Build a Predictive Network--based Walkability Scoring System. \textit{PLoS
  One} \textbf{17}(6), e0270098 (2022).

\bibitem{papadakis2019function}
E.~Papadakis, G.~Baryannis, A.~Petutschnig, and T.~Blaschke, Function-based
  Search of Place Using Theoretical, Empirical and Probabilistic Patterns.
  \textit{ISPRS International Journal of Geo-Information} \textbf{8}(2), 92
  (2019).

\bibitem{barbosa2021uncovering}
H.~Barbosa, S.~Hazarie, B.~Dickinson, A.~Bassolas, A.~Frank, H.~Kautz,
  A.~Sadilek, J.~J. Ramasco, and G.~Ghoshal, Uncovering the Socioeconomic
  Facets of Human Mobility. \textit{Scientific Reports} \textbf{11}(1), 8616
  (2021).

\bibitem{JLL}
I.~Jones Lang LaSalle~IP, Use our interactive tool to compare 300 cities across
  a range of economic and real estate indicators.
  \url{https://www.us.jll.com/en/trends-and-insights/research/global/crc/city-sunburst-chart}
  (2018). (Accessed on 02/01/2023).

\bibitem{mimar2022connecting}
S.~Mimar, D.~Soriano-Pa{\~n}os, A.~Kirkley, H.~Barbosa, A.~Sadilek, A.~Arenas,
  J.~G{\'o}mez-Garde{\~n}es, and G.~Ghoshal, Connecting intercity mobility with
  urban welfare. \textit{PNAS Nexus} \textbf{1}(4), pgac178 (2022).

\bibitem{IMF}
{International Monetary Fund}, {World Economic and Financial Surveys World
  Economic Outlook Database---WEO Groups and Aggregates Information}.
  \url{https://www.imf.org/en/Publications/SPROLLs/world-economic-outlook-databases}
  (2023).

\bibitem{wiki}
 (2022), \urlprefix\url{https://wiki.openstreetmap.org/wiki/Main_Page}.

\bibitem{LOSI}
City Data.
  \url{https://publicadministration.un.org/egovkb/en-us/Data/City?fbclid=IwAR36nph_BSYSIVz_ABzLc9N4xpSsbEamreG7NsCTiyaFLt_A6NxpurY7P00}.
  (Accessed on 02/01/2023).

\bibitem{WalkScore}
Most Walkable Cities in the United States and Canada on Walk Score.
  \url{https://www.walkscore.com/cities-and-neighborhoods/}. (Accessed on
  02/01/2023).

\bibitem{carr2010walk}
L.~J. Carr, S.~I. Dunsiger, and B.~H. Marcus, Walk Score{\textregistered} as a
  global estimate of neighborhood walkability. \textit{American Journal of
  Preventive Medicine} \textbf{39}(5), 460--463 (2010).

\bibitem{duncan2011validation}
D.~T. Duncan, J.~Aldstadt, J.~Whalen, S.~J. Melly, and S.~L. Gortmaker,
  Validation of Walk Score{\textregistered} for estimating neighborhood
  walkability: an analysis of four US metropolitan areas. \textit{International
  Journal of Environmental Research and Public Health} \textbf{8}(11),
  4160--4179 (2011).

\bibitem{dijkstra2022note}
E.~W. Dijkstra, A Note on Two Problems in Connecion with graphs. In
  \textit{Edsger Wybe Dijkstra: His Life, Work, and Legacy}, pp. 287--290
  (2022).

\bibitem{cooper2020sdna}
C.~H. Cooper and A.~J. Chiaradia, sDNA: 3-d Spatial Network Analysis for GIS,
  CAD, Command Line \& Python. \textit{SoftwareX} \textbf{12}, 100525 (2020).

\bibitem{hanna2021random}
S.~Hanna, Random Walks in Urban Graphs: A Minimal Model of Movement.
  \textit{Environment and Planning B: Urban Analytics and City Science}
  \textbf{48}(6), 1697--1711 (2021).

\bibitem{kirkley2018betweenness}
A.~Kirkley, H.~Barbosa, M.~Barthelemy, and G.~Ghoshal, From the Betweenness
  Centrality in Street Networks to Structural Invariants in Random Planar
  Graphs. \textit{Nature Communications} \textbf{9}(1), 2501 (2018).

\bibitem{yang2018universal}
H.~Yang, J.~Ke, and J.~Ye, A Universal Distribution Law of Network Detour
  Ratios. \textit{Transportation Research Part C: Emerging Technologies}
  \textbf{96}, 22--37 (2018).

\bibitem{prato2009route}
C.~G. Prato, Route Choice Modeling: Past, Present and Future Research
  Directions. \textit{Journal of Choice Modelling} \textbf{2}(1), 65--100
  (2009).

\bibitem{barbosa2018human}
H.~Barbosa, M.~Barthelemy, G.~Ghoshal, C.~R. James, M.~Lenormand, T.~Louail,
  R.~Menezes, J.~J. Ramasco, F.~Simini, and M.~Tomasini, Human Mobility: Models
  and Applications. \textit{Physics Reports} \textbf{734}, 1--74 (2018).

\bibitem{rosenthal2004evidence}
S.~S. Rosenthal and W.~C. Strange, Evidence on the Nature and Sources of
  Agglomeration Economies. In \textit{Handbook of Regional and Urban
  Economics}, volume~4, pp. 2119--2171, Elsevier (2004).

\bibitem{walkscoremethodology}
W.~Score, Walk Score Methodology. \url{https://www.cedeus.cl/} (2011).
  (Accessed on 02/01/2023).

\bibitem{hall2018walk}
C.~M. Hall and Y.~Ram, Walk score{\textregistered} and its potential
  contribution to the study of active transport and walkability: A critical and
  systematic review. \textit{Transportation Research Part D: Transport and
  Environment} \textbf{61}, 310--324 (2018).

\bibitem{jonckheere1954distribution}
A.~R. Jonckheere, A distribution-free k-sample test against ordered
  alternatives. \textit{Biometrika} \textbf{41}(1/2), 133--145 (1954).

\bibitem{UnHabitat}
Datasets | Urban Indicators Database.
  \url{https://data.unhabitat.org/pages/datasets}. (Accessed on 02/01/2023).

\bibitem{huang2015circuity}
J.~Huang and D.~M. Levinson, Circuity in urban transit networks.
  \textit{Journal of Transport Geography} \textbf{48}, 145--153 (2015).

\end{thebibliography}
\end{document}